\documentclass[natbib]{svjour3}                     
\smartqed  
\usepackage{graphicx}
 \usepackage{mathptmx}      
 \usepackage{aps-bibstyle}  
\usepackage{latexsym, amsfonts, mathcomp}
\usepackage{amssymb}
\newcommand{\aap}{{Astron. Astrophys.}}
\newcommand{\apj}{{Astrophys. J.}}

\newcommand{\apjl}{{Astrophys. J. Lett.}}
\newcommand{\mnras}{"Monthly Notices of the Royal Ast. Society"}
\newcommand{\nat}{"Nature"}

\def\approxgt{\ifmmode \rlap{$>$}{}_{{}_{{}_{\textstyle\sim}}} \else%
$\rlap{$>$}{}_{{}_{{}_{\textstyle\sim}}}$\fi} 
\def\approxlt{\ifmmode \rlap{$<$}{}_{{}_{{}_{\textstyle\sim}}} \else%
$\rlap{$<$}{}_{{}_{{}_{\textstyle\sim}}}$\fi}

\def\xmm{XMM-{\it Newton}}
\def\chan{{\it Chandra}}

\begin{document}

\title{Mass Measurements of Stellar and Intermediate-Mass Black-Holes 
}

\titlerunning{Black hole mass measurements}        

\author{J. Casares         \and
        P.G. Jonker 
}


\institute{J. Casares \at
 Instituto de Astrof\'isica de Canarias,  E--38205 La Laguna, S/C de Tenerife, Spain\\
Departamento de Astrof\'isica, Universidad de La Laguna, E--38206 La Laguna, S/C de Tenerife, Spain
              \email{jorge.casares@iac.es}   
           \and
           P.G. Jonker \at
              SRON, Netherlands Institute for Space Research, Sorbonnelaan 2,
  3584~CA, Utrecht, The Netherlands\\
  Department of Astrophysics/IMAPP, Radboud University Nijmegen,
P.O.~Box 9010, 6500 GL, Nijmegen, The Netherlands\\
Harvard--Smithsonian  Center for Astrophysics, 60 Garden Street, Cambridge, MA~02138, U.S.A.
}

\date{Received: date / Accepted: date}

\maketitle

\begin{abstract}
  We discuss the method, and potential systematic effects therein, used for
  measuring the mass of stellar-mass black holes in X-ray binaries. We
  restrict our discussion to the method that relies on the validity of
  Kepler's laws; we refer to this method as the dynamical method. We
  briefly discuss the implications of the mass distribution of
  stellar-mass black holes and provide an outlook for future
  measurements. Further, we investigate the evidence for the existence
  of intermediate-mass black holes i.e.  black holes with masses
  above 100 M$_\odot$, the limit to the black hole mass that can be
  produced by stellar evolution in the current Universe. 
\end{abstract}

\section{Introduction}
\label{sec:1}

Black holes (BH) are the pinnacle of extreme gravity. They provide
astronomers with unique laboratories for observing some of the most
fundamental and intriguing astrophysical phenomena, such as accretion,
the ejection of relativistic outflows or the production of gamma-ray
bursts. Consequently, BHs play an essential role in a variety of 
astrophysical phenomena on various scales, ranging  from binaries to
ultra-luminous X-ray sources (ULXs) , galaxies and quasars, the most
powerful accretion engines in the Universe. However, it is {\it
  stellar-mass} BHs that offer us the best opportunity to study these
objects in detail.  Their proximity and variability time scales allow
in-depth studies of their properties through a range of accretion
regimes on time-scales convenient for study by humans.  Thorough
reviews on BH accretion and outflows are given in 
several other articles of this issue.

Astrophysical BHs are characterized by only two parameters, mass and spin,
and their knowledge is key to probe space-time in the strong gravity 
regime (see articles by McClintock et al. and Reynolds in this issue). 
Accurate knowledge of BH masses is also critical to test  
models of massive progenitors, SNe Ibc explosions and compact binary evolution 
(e.g. \citealt{fryer01, fryer12, belczynski12}). The current article presents 
an up-to-date overview of dynamical mass determinations in stellar-mass BHs. 
The main methods of analysis are summarized, together with a critical 
assessment on their limitations and possible systematics involved. In a second 
part of the article, new approaches and techniques are reviewed from 
which a significant advance in the precision of mass measurements is expected. 
Finally, a section on prospects for mass determination in ULXs 
is presented. Previous reviews on 
observational properties and mass determination in BH binaries 
can be found in 
\cite{vanparadijs95}, \cite{tanaka96}, 
\cite{orosz03}, \cite{charles06}, 
\cite{remillard06}, \cite{mcclintock06}, 
\cite{casares07} and \cite{belloni11}.

\section{Dynamical BHs in X-ray Transients}
\label{sec:2}

Stellar evolution predicts $\gtrsim10^8$ BH remnants in the Galaxy
\citep{vandenheuvel92}  
but only BHs in compact binaries can be easily detected
through accretion.  X-ray binaries thus provide currently the best way
to measure the mass of BHs. A large number of these X-ray binaries are 
found as X-ray transients (XRTs, for a comprehensive review see 
\citealt{mcclintock06}). XRTs are singled out by episodic 
outbursts caused by mass transfer instabilities in an accretion disc
which is fed by a low-mass (donor) star 
\citep{mineshige89,lasota01}.  The large fraction of BH systems 
among XRTs agrees with the predictions of the Disc Instability Model,
modified by irradiation effects. The absence of a solid surface in
the compact star inhibits disc stabilization through X-ray irradiation
at the low accretion rates characteristic of overflowing low-mass
stars \citep{king97,coriat12}. XRTs may increase the integrated X-ray
luminosity of the Milky Way by a factor $\sim$2 and thus are promptly
spotted by X-ray satellites. Between outbursts, they
remain in a "quiescent" state, with typical X-ray luminosities below
$\sim10^{32}$ erg s$^{-1}$, allowing the optical detection of the
faint low-mass donor star.  This opens-up the possibility to perform
radial velocity studies, probe the nature of the compact star and
determine its mass. 

\begin{figure}
  \includegraphics[width=0.45\textwidth,angle=-90]{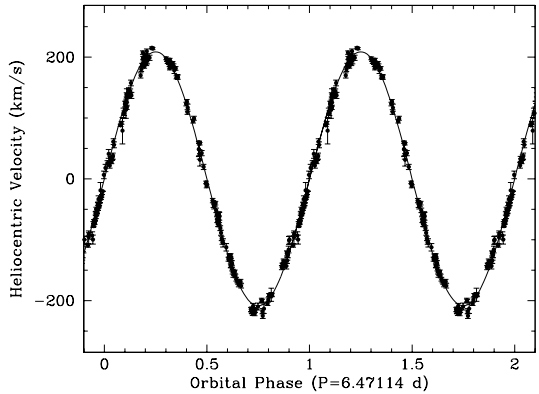}
\caption{Radial velocity curve of the K0 donor star 
in the XRT V404 Cyg.}
\label{fig:1}       
\vspace*{3em}
\end{figure}

The most robust method of measuring stellar masses relies on Kepler's 
Third law of motion. 
However, BH X-ray binaries are akin to single-lined spectroscopic binaries and 
hence only the radial velocity curve of the mass-losing star is available 
(Fig.~\ref{fig:1}). This yields
the orbital period $P_{\rm orb}$ and the radial velocity semi-amplitude of the 
companion star $K_{\rm c}$. The two quantities combine in the mass function 
equation $f(M)$, a non-linear expression relating the masses of the compact 
object $M_{\rm x}$ and the companion star $M_{\rm c}$ (or binary mass ratio 
$q=M_{\rm c}/M_{\rm x}$) with the binary inclination angle ($i$): 

\begin{equation}
f(M) =  {K_{\rm c}^{3} P_{\rm orb}\over{2 \pi G}} = 
{M^{3}_{\rm x}\sin^{3} i\over{\left(M_{\rm x} + M_{\rm c}\right)^2}} = 
{M_{\rm x}\sin^{3} i\over{(1+q)^2}}
\end{equation}
  
\noindent
Note that this expression implicitly assumes orbits are circular, 
which is a fair assumption given the long lifetimes and short circularization 
timescales expected in X-ray binaries \citep{witte01}. 
The mass function provides a solid lower limit to the mass of the
compact star for $M_{\rm c}=0$ and an edge-on geometry
($i=90^{\circ}$). A large mass function is widely considered as the
best signature for a BH since observations and theoretical
calculations of dense matter indicate that neutron stars cannot be
more massive than $\sim$2.5 M$_{\odot}$ \citep{lattimer12}. 
Besides, main sequence stars with masses above $\sim$3 M$_{\odot}$ would 
be B-type stars, and would be seen in the spectra. The fact that only a 
K/M-type stellar spectrum is detected together with the presence of an
occasionally very bright X-ray 
source leads to the conclusion that the binary contains a BH.

The radial velocity curve of the donor star is best obtained through
cross-correlation of the photospheric absorption lines with a stellar
template of similar spectral type. Mass functions are gathered
routinely to a few percent accuracy and this requires resolving powers
better than about $\lambda/\Delta\lambda$=1500. The use of 10~m class
telescopes has allowed one to measure $f(M)$ for objects down to R$\sim$22
mag, as shown by the works on XTE J1859+226
\citep{filippenko01, corral11}.  BH mass measurements require also the
measurement of the mass ratio $q$ and the binary inclination which, in
the absence of eclipses, can only be obtained through indirect
methods. These are based on information to be extracted from the {\it
  light curve} and the {\it spectrum} of the optical star, resulting
in a full solution to the binary parameters with minimum
assumptions. This procedure will be discussed in turn.

\begin{figure}
  \includegraphics[width=0.45\textwidth,angle=-90]{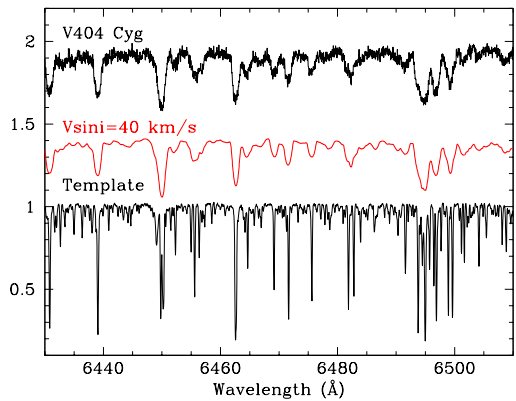}
\caption{Rotational broadening analysis. 
A K0IV template star (bottom) is broadened by $V \sin i=$40 km s$^{-1}$ (middle) 
in order to reproduce the spectrum of the donor star in the BH transient V404 
Cyg (top). The latter has been produced after coadding individual spectra in 
the rest frame of the companion star.}
\label{fig:2}       
\end{figure}
 
The best way to determine the mass ratio is through measuring the rotational 
broadening ($V \sin i$) of the photospheric lines from the companion star. This 
technique exploits the fact that the star fills its Roche lobe and it is 
tidally locked. This makes the absorption lines significantly broader than 
in single stars that are slowly rotating.
Under the approximation of sphericity, the rotational 
broadening scales with the velocity of the donor star according to               
$V \sin i/K_{\rm c}\simeq 0.462~q^{(1/3)} (1+q)^{(2/3)}$ \citep{wade88} and 
hence $q$ can be constrained. The rotational broadening is usually measured by 
comparing the target spectrum with a slowly rotating template, convolved with 
a limb-darkened rotational profile (e.g. \citealt{gray92}). A $\chi^2$ 
minimization yields the optimum broadening required by the template to match 
the spectrum of the companion to the BH (see Fig.~\ref{fig:2}). Typical 
rotational broadenings in BH transients range between 30 and 120 km s$^{-1}$ 
and thereby moderately high spectral resolutions 
($\lambda/\Delta\lambda\gtrsim5000$) are required for these measurements
otherwise erroneous values for $V \sin i$ are likely obtained.  

It should be noted that there are systematic errors involved in the calculation 
of $V \sin i$. First of all, Roche-lobe-filling stars are obviously 
non-spherical, with tidal distortion causing orbital variations of the 
broadening kernel. Fitting the phase-averaged spectrum with a 
template broadened using a spherical convolution profile yields 
$V \sin i$ values which underestimate $q$ \citep{marsh94}.  
In addition, the use of a continuum limb-darkening approximation 
also leads to underestimates of the true broadening and thus a decreased mass 
ratio~\citep{shahbaz03a}. Another potential source of systematics is introduced 
by assuming a gravity darkening 
law described by von Zeipel's theorem with exponent $\beta=0.08$ 
(\citealt{lucy67}, but see \citealt{sarna89}).
In any case, statistical uncertainties in the computation of $V \sin i$ are 
typically larger than systematic errors and hence mass ratios obtained through 
measuring rotational broadenings are, in most cases, robust. Furthermore, given 
the extreme mass ratios ($q \ll 1$) typical of BH XRTs, the impact of $q$ 
uncertainties in the final BH mass is modest.   

\begin{figure}
  \includegraphics[width=0.6\textwidth,angle=-90]{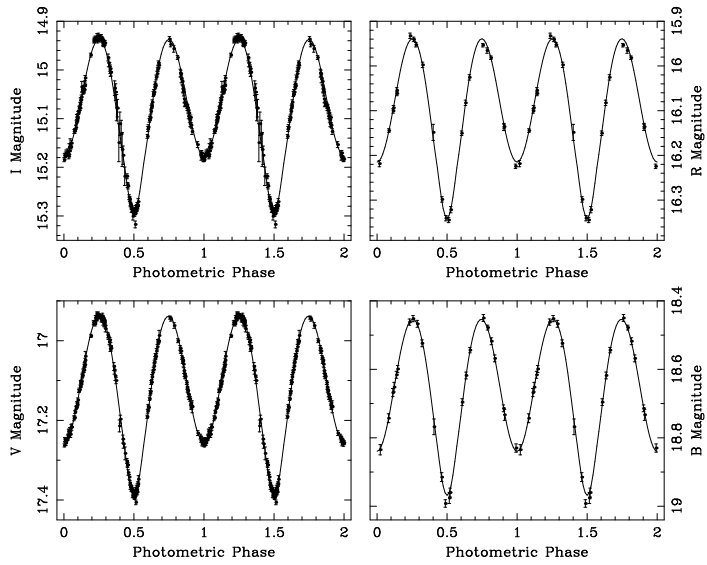}
\caption{Ellipsoidal fits to light curves of the 
XRT GRO J1655-40 in four colour bands simultaneously. Synthetic models were 
computed using Kurucz model atmospheres. From \cite{beer02}. }
\label{fig:3}       
\end{figure}

The binary inclination is commonly obtained through fitting 
optical/NIR light curves with synthetic ellipsoidal models. Light curves in 
XRTs show 
a characteristic double-humped modulation produced by the tidal distortion of
the Roche lobe filling
donor star and a non-uniform distribution of surface brightness. 
The amplitude of the modulation is a strong function of the inclination angle. 
Synthetic models are computed integrating the local flux intensity, modified by 
limb and gravity darkening effects, over the Roche geometry. The best results 
are obtained using Kurucz and NEXTGEN model atmospheres (see \citealt{orosz00}
for a critical comparison of several approaches). 
For example, Fig.~\ref{fig:3} presents 
a textbook example of the ellipsoidal modulation from GRO J1655-40, an XRT with 
a F6IV intermediate-mass donor star. Synthetic model fits performed by 
different groups have resulted in very accurate inclination values distributed 
over a narrow range between 64-71$^{\circ}$ \citep{orosz97,vanderhooft98,greene01,beer02}. 
However, the vast majority of XRTs possess faint 
K-M donor stars and, 
therefore, light curves can be seriously contaminated by other non-stellar 
sources of light. 
The impact of these on the determination of 
inclination angles and BH masses can be critical and will 
be discussed in Section ~\ref{sec:2.1}. 

Table~\ref{tab:1} presents a compilation of fundamental parameters and
BH mass determinations for the 17 BH XRTs with dynamical confirmation
currently known. Uncertainties are typically 1-$\sigma$ except for
errors in the inclination angle where a 90 or 95 percent confidence
level is sometimes provided. The mass functions listed in the table
have been obtained from radial velocity curves of the companion stars
in quiescence, with the exception of GX 339-4 (see also footnotes on
GRS 1915+105 and GRO J1655-40). In the case of GX 339-4, the donor
star has not been detected in quiescence yet, although a lower limit
to the mass function was derived using emission lines, excited on its
irradiated hemisphere during an outburst episode (\citealt{hynes03a};
see Section 4.1).  Table~\ref{tab:1} also quotes mass ratios obtained
exclusively through the $V \sin i$ technique, besides that from GRO
J1655-40 where its high mass ratio also allows one to constrain $q$ from
model fits to the ellipsoidal light curves. Finally, binary inclinations refer to
values derived from modelling ellipsoidal light curves in quiescence,
except for GRS 1915+105, where it has been inferred from the
orientation of the radio jets \citep{fender99}.  In some cases, upper
limits to the inclinations are given, based on the lack of X-ray
eclipses and mass ratio constraints.  Numbers highlighted in
boldface indicate our recommended set of fundamental parameters to
be adopted. In our view, these provide the best
determinations currently available i.e.~those least affected by possible
systematic effects and having the lowest statistical uncertainties.

\subsection{Systematic errors and biases in BH mass determinations}
\label{sec:2.1}

\begin{figure}
  \includegraphics[width=0.90\textwidth,angle=0]{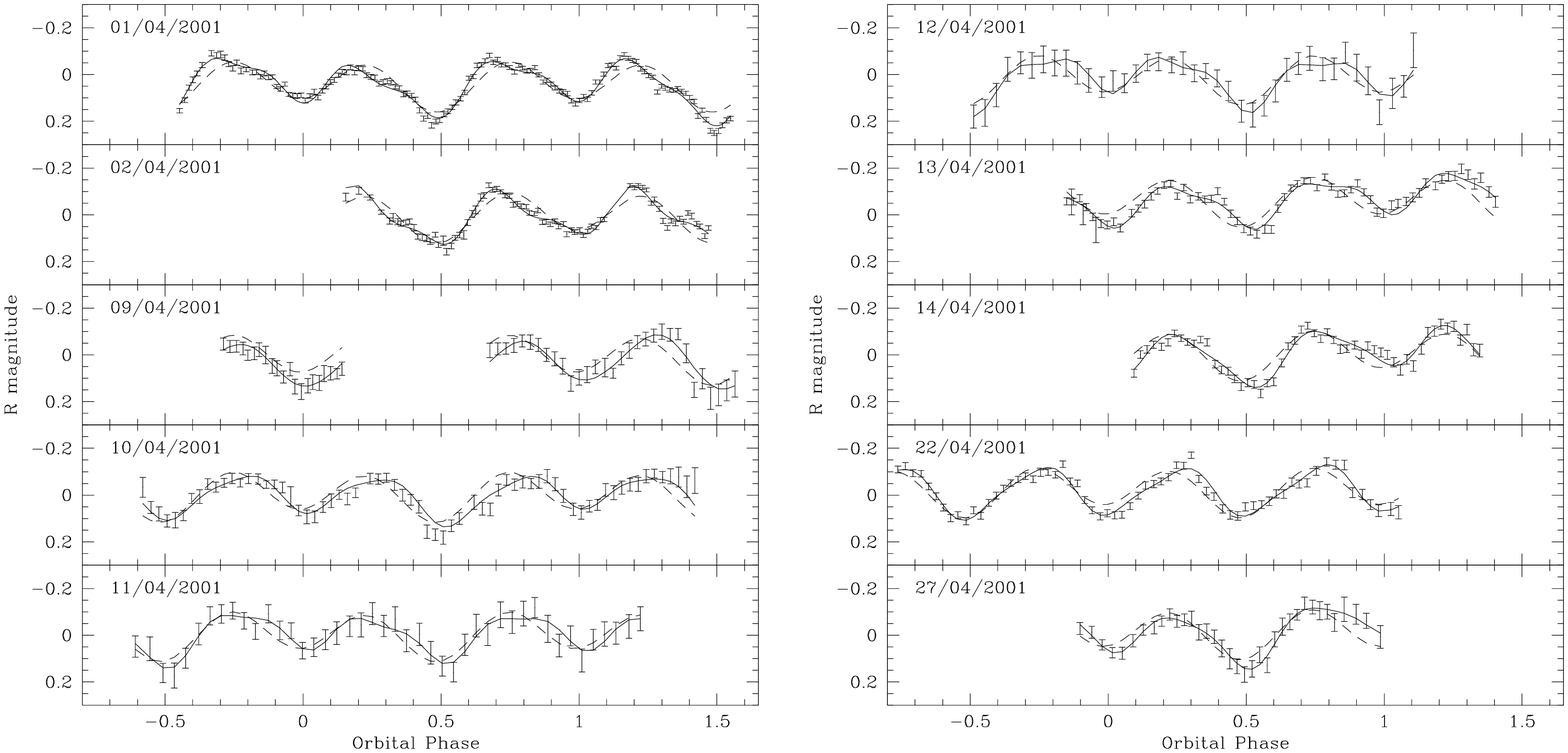}
\caption{Ellipsoidal light curves distorted by a
superhump in XTE J1118+480. A combined model of an ellipsoidal plus superhump
waves (continuous line) provides a better description of the data than a pure
ellipsoidal fit (dashed line). From \cite{zurita02}.}
\label{fig:4}       
\end{figure}

The error bars on BH mass measurements are dominated by uncertainties
in the inclination angle because of its cubic dependence in equation
1.  But most worryingly, Table~\ref{tab:1} indicates that ellipsoidal
fits performed by independent groups on the same binary often lead to
a wide spread of inclinations and thereby BH masses. This is mainly
thought to be due to systematic effects caused by contamination from
non-stellar sources of light rather than statistical errors. There are
two main sources of systematics 
affecting light curve analysis. 
The first one is the presence of a {\it superhump}
modulation, a distorting wave produced by an eccentric disc precessing
with a timescale a few percent longer than the orbital
period. Superhumps are typically seen in outburst, when the accretion
disc exceeds the 3:1 resonance radius \citep{odonoghue96}, but can
also be detected near quiescence. When this happens, the ellipsoidal
light curve is distorted by secular changes in shape and relative
height of the maxima and minima (Fig.~\ref{fig:4}). Intensive
monitoring over several orbital cycles is thus important to
disentangle potential superhump waves from the true ellipsoidal
modulation. Sometimes asymmetries in ellipsoidal light curves are
interpreted as contamination by a hot spot (when extra flux is located
at phase $\sim0.75$, e.g. \citealt{khargharia13}) or stellar spots.
Observations of sharp asymmetries in the light curves can also be
mistaken for eclipse features, leading to overestimates in the
inclination angle \citep{haswell93}.

\begin{figure}
  \includegraphics[width=0.50\textwidth,angle=0]{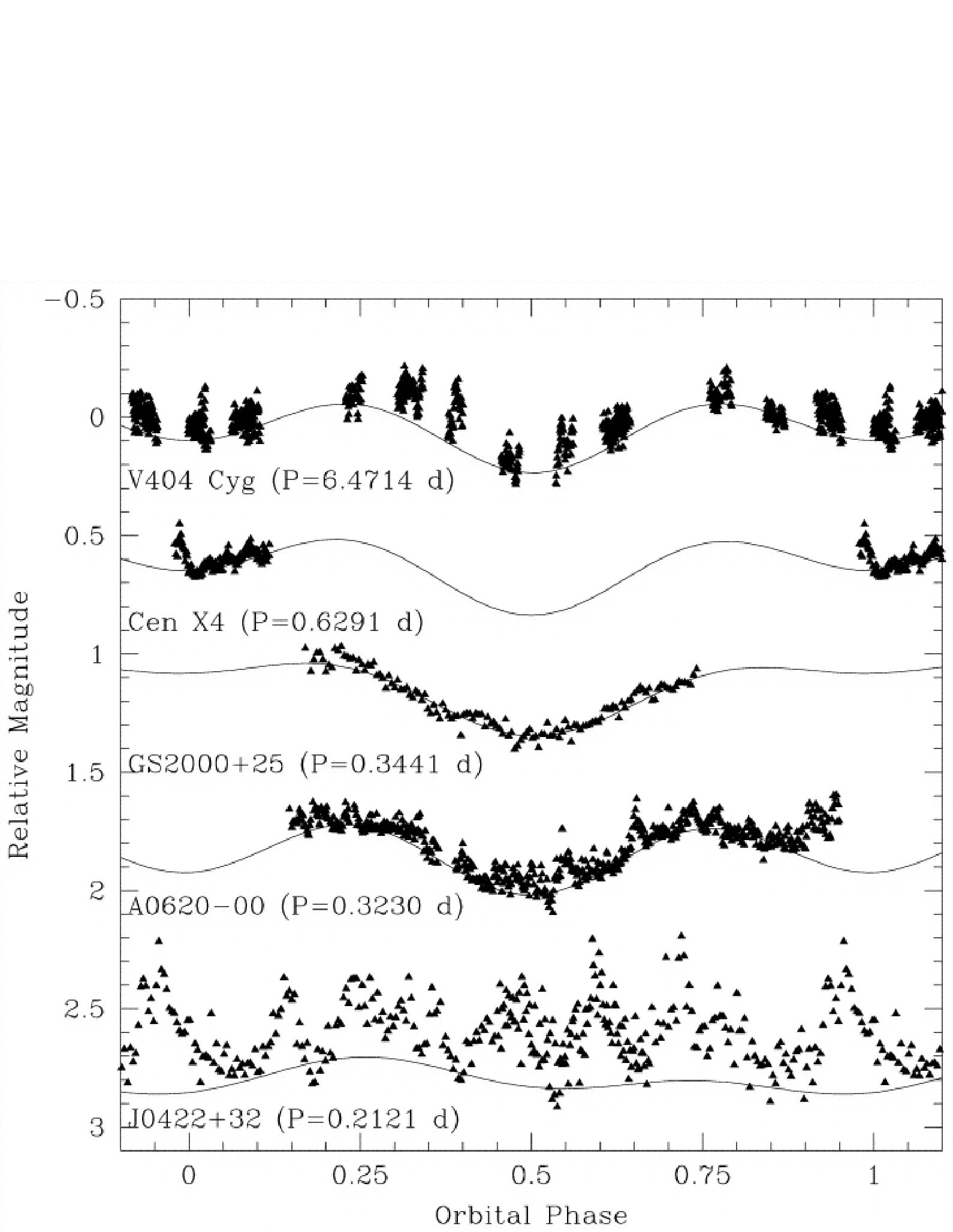}
\caption{High-time resolution light curves of 
quiescent XRTs showing the presence of fast variability. From \cite{zurita03}.}
\label{fig:5}      
\end{figure}

The second source of systematics is caused by contamination from rapid
aperiodic variability. The presence of optical flares in V404 Cyg,
with a timescale of $\sim$6hr, was already noticed long ago but
thought to be peculiar to this system
\citep{wagner92,pavlenko96}. However, subsequent high-time resolution
(1-5 mins) light curves have revealed that all quiescent XRTs display
the same type of variability, with typical amplitudes ranging from
0.06 to 0.6 mag (\citealt{zurita03,hynes03b}; see Fig.~\ref{fig:5} but
also \citealt{shahbaz13} for a record $\sim$1.5 mag amplitude
flaring activity). The variability seems stronger for systems with cooler
companions and its characteristic time-scale appears to increase with
orbital period, both properties suggesting an accretion disc
origin. Magnetic reconnection events \citep{zurita03}, X-ray
reprocessing \citep{hynes04}, instabilities in the transition between
the thin and advective disc \citep{shahbaz03b} and variable synchrotron
emission from a disc jet/outflow \citep{shahbaz13} have been proposed
but the physical mechanism responsible for the rapid variability is
still unknown.

It is commonly assumed that the accretion disc light follows a negative power 
law with $\lambda$ (see Table 1 in \citealt{garcia96}) and thereby most ellipsoidal fits 
tend to be performed at NIR wavelengths to minimize contamination. 
However, strong flaring activity is also observed in several NIR light curves, questioning 
this strategy. The most dramatic case is presented by K-band observations of GRO J0422+320,
where the ellipsoidal modulation becomes completely diluted by the flaring variability 
\citep{reynolds07}. 
Further, flickering is not a white noise process (the PDS is described by a negative 
power-law with index $\sim$-1.3, see \citealt{shahbaz03b}) and thus it is
not cancelled out after binning or averaging light curves over many orbital cycles. 
The aperiodic variability can also vary with time for any particular
system as shown by several cases in the literature (e.g. BW Cir: \citealt{casares09};  
XTE J1859+226: \citealt{corral11}). 

\begin{figure}
  \includegraphics[width=0.70\textwidth,angle=0]{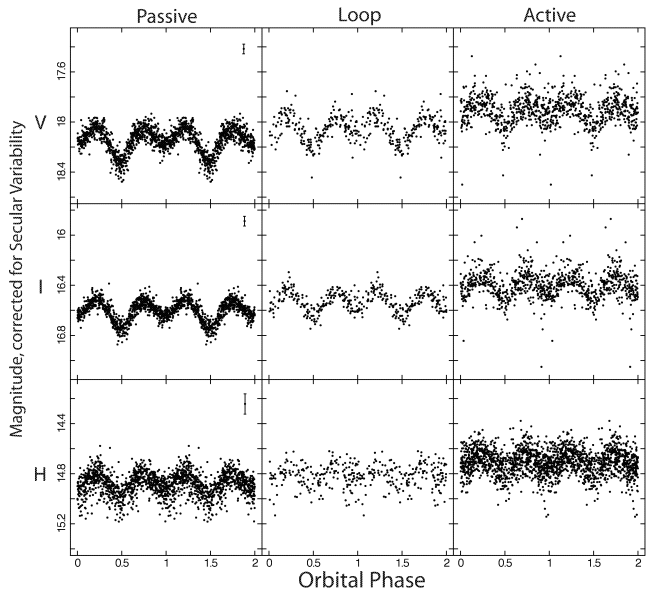}
\caption{Ellipsoidal light curves of A0620-00 in three different 
quiescent states: {\it passive} (left panels), {\it loop} (middle panels) and {\it active} (right).  
The strength of the aperiodic variability increases from the {\it passive} to the {\it active} 
state. From \cite{cantrell08}.}
\label{fig:6}       
\end{figure}

Using a decade-long data set of multicolour quiescent photometry, 
\cite{cantrell08} identify two main states of variability in A0620-00. The so-called 
{\it passive} and {\it active} states fall in separate parts of a colour-magnitude 
diagram. The system becomes redder in the {\it passive} state and displays minimum flaring 
activity (Fig. \ref{fig:6}). Ellipsoidal fits to a subset of {\it passive} $VIH$ 
light curves separately yield an inclination which is $\sim$10$^{\circ}$ 
higher than previous works have suggested \citep{cantrell10}. The latter were based on 
pure ellipsoidal fits to NIR light curves where the non-stellar contribution was neglected. 
Consequently, the mass of the BH in A0620-00 can be overestimated by a factor $\approx$2 if 
the disc contamination is ignored. 
This work has demonstrated that it is critial to employ light curves with 
minimum flickering activity during {\it passive} states to measure unbiased 
binary inclinations. 
Further discussion on the impact of the rapid variability in mass determinations of other 
BH XRTs is presented in \cite{kreidberg12}.  

\begin{figure}
  \includegraphics[width=0.7\textwidth,angle=0]{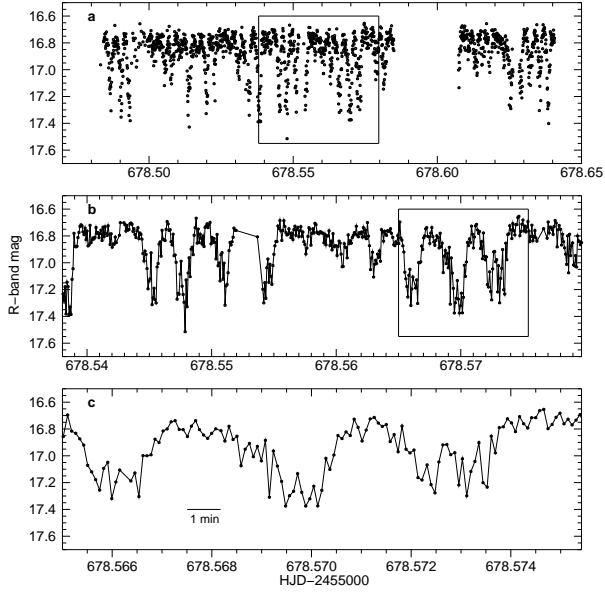}
\caption{Optical light curves of Swift\,J1357.2$-$0933 
during outburst at 7\,s time resolution. Regular dips repeating every 2\,min cause 
a drop in brightness of up to $\simeq0.8$\,mag. Close ups with lengths of 1\,h and 
15\,min are displayed in the middle and bottom panels, respectively. 
From \cite{corral13}.}
\label{fig:7}       
\end{figure}

One of the peculiar properties of the sample of BH XRTs listed in
Table~\ref{tab:1} is the absence of binaries with inclinations
$i>75^{\circ}$.  Furthermore, none of the other $\sim$33 BH candidates
(i.e. those XRTs with similar X-ray properties to dynamical BHs) shows
eclipses, whereas $\sim$20 percent are expected for a random
distribution of inclinations (although this number is dependent on the
assumed mass ratio, $q$). The lack of eclipsing BH XRTs is intriguing
and strongly suggests that an observational bias is at play.  It has
been proposed that high inclination systems are hidden from view due
to obscuration of the central X-ray source by a flared accretion disc
\citep{narayan05}. The recent discovery of optical dips in the XRT
Swift\,J1357.2$-$0933 suggests that the first extreme inclination BH
transient may have been detected (Fig.~\ref{fig:7}). Aside from the
unusual optical dips, the system is remarkable because of its
extremely broad H$_{\alpha}$ emission profile and very low peak X-ray
luminosity, properties which can be explained by orientation effects
in an edge-on geometry (\citealt{corral13}; but see Armas Padilla 2013a, 2013b 
for a different interpretation which proposes an intrinsically faint XRT). 
Based on the double-peak separation and radial velocities of the H$_{\alpha}$ 
profile indirect prove for a BH in a 2.8 hr orbit is presented. In addition, 
evidence is provided for the presence of an obscuring torus in the inner disc.
This brings a new ingredient to theoretical modelling of inner disc
flows and jet collimation mechanisms in stellar-mass BHs. The
discovery of edge-on BH XRTs is important because these systems will
likely deliver the most precise BH mass determinations.  Therefore,
they will be key in the construction of the mass distribution of
compact remnants.
  
\begin{table}
\caption{Mass measurements of BHs in XRTs (continues on next page). We
  mark in bold face the measurements we consider the most reliable
  currently available in the literature. 
}
\label{tab:1}       
%
%

\begin{tabular}{lccccccc}
\hline\hline\noalign{\smallskip}
 System & Donor  &  $P_{\rm{}orb}$ &  $f(M)$ & q & i & $M_{\rm x}$ & Ref.\\
 &  Spect. Type & [days] &  [$M_{\odot}$] &  & [deg] &  [$M_{\odot}$] & \\ 
\noalign{\smallskip}\hline\hline\noalign{\smallskip}
GRS 1915+105 & K1/5 III & 33.5(1.5) & 9.5$\pm$3.0& 0.058$\pm$0.033       & 66$\pm$2       & 14.0$\pm$4.4        &[1-2]\\
~~~~~~~,,        &           & 33.8(1)      & 8.0$\pm$0.6     &                       & 66$\pm$2       & 12.0$\pm$1.4        &[3]\\
~~~~~~~,,   &   & {\bf 33.85(16)} & {\bf 7.02$\pm$0.17} & {\bf 0.042$\pm$0.024}  & {\bf 66$\pm$2} & {\bf 10.1$\pm$0.6}        &[4]\\
\noalign{\smallskip}\hline\noalign{\smallskip}
V404 Cyg        &  K0 IV     & {\bf 6.47129(7)}  & {\bf 6.07$\pm$0.05}   & {\bf 0.067$\pm$0.005}       &                &                     &[5-7]\\
~~~~~~~,,       &            &              &                 &                       & 60$-$80        & 8$-$12              &[8]\\
~~~~~~~,,       &            &              &                 &                       & 56$\pm$4       & 12$^{+3}_{-2}$      &[9]\\
~~~~~~~,,       &            &              &                 &                       & 56$\pm$2       &                     &[10]\\
~~~~~~~,,       &            &              &                 &                       &  $>$62         & $\lesssim$12.5      &[11]\\
~~~~~~~,,       &  K3 IV     &              &                 &                       & 67$^{+3}_{-1}$ & 9.0$^{+0.2}_{-0.6}$ &[12]\\
\noalign{\smallskip}\hline\noalign{\smallskip}
XTE J1819.3-2525&  B9 III    & {\bf 2.81730(1)}   &  {\bf 2.74$\pm$0.12}  & {\bf 0.63$-$0.70}   & 60$-$71        &  8.7$-$11.7         &[13]\\
\noalign{\smallskip}\hline\noalign{\smallskip}
GRO J1655-40    & $\sim$F5 IV & 2.601(27)   &  3.16$\pm$0.15  &                       &                &                     &[14]\\
~~~~~~~,,	     &  F3/6 IV   & 2.62157(15) & 3.24$\pm$0.09 & 0.33$\pm 0.01$      & 69.5$\pm$0.1   &  7.0$\pm$0.2        &[15]\\
~~~~~~~,,	     &  	  & {\bf 2.62168(14)} &         & 0.24$-$0.41         & 63.7$-$70.7    &  6.3$-$7.6          &[16]\\
~~~~~~~,,	     &  	  & 2.62191(20) &               & 0.34$-$0.4          & 70.2$\pm$1.9   &  6.3$\pm$0.5        &[17]\\
~~~~~~~,,	     &  F5/G0 IV  &             &               & 0.26$\pm 0.04$      & {\bf 68.7$\pm$1.5}  & 5.4$\pm$0.3  &[18]\\
~~~~~~~,,	     &  F6 IV	  &    & {\bf 2.73$\pm$0.09} & 0.34$-$0.44         &                &  5.5$-$7.9          &[19]\\
~~~~~~~,,	     &  	  &		&               & 0.31$\pm 0.08$      &                &  7.9$\pm$3.8        &[20]\\
~~~~~~~,,	     &  	  &		&               & {\bf 0.42$\pm$0.03}  &              &  {\bf 6.6$\pm$0.5}        &[21]\\
~~~~~~~,,	     &  	  & 2.62120(14)	& 3.16$\pm$0.15 & 0.33$\pm 0.05$      &                &                     &[22]\\
\noalign{\smallskip}\hline\noalign{\smallskip}
\end{tabular}
\end{table}

%
%

\noindent
{\small 
\begin{tabular}{lccccccc}
\hline\hline\noalign{\smallskip}
 System & Donor  &  $P_{\rm{}orb}$ &  $f(M)$ & q & i & $M_{\rm x}$ & Ref.\\
 &  Spect. Type & [days] &  [$M_{\odot}$] &  & [deg] &  [$M_{\odot}$] & \\ 
\noalign{\smallskip}\hline\noalign{\smallskip}
BW Cir               &$\sim$G5 IV & {\bf 2.54451(8)}  & {\bf 5.73$\pm$0.29} & {\bf 0.12$^{+0.03}_{-0.04}$}& $<$79        &  $>$7.0    &[23-24]\\
\noalign{\smallskip}\hline\noalign{\smallskip}
GX 339-4 &  -- & {\bf 1.7557(4)} & {\bf 5.8$\pm$0.5}   &                 &                & $>$6.0      &[25-26]\\
\noalign{\smallskip}\hline\noalign{\smallskip}
XTE J1550-564   &  K2/4  IV  & {\bf 1.5420333(24)}&  {\bf 7.65$\pm$0.38}  & $\approx$0.03       & 74.7$\pm$3.8   &  7.8$-$15.6       &[27-28]\\
\noalign{\smallskip}\hline\noalign{\smallskip}
4U 1543-475     &  A2 V	     & {\bf 1.123(8)}     &  {\bf 0.22$\pm$0.02}    &                     & 20$-$40        &  2.7$-$7.5          &[29]\\
\noalign{\smallskip}\hline\noalign{\smallskip}
H1705-250       &  K3/M0 V   & {\bf 0.5213(13)} &  {\bf 4.86$\pm$0.13}  &  $\le$0.053   & 60$-$80        &  4.9$-$7.9          &[30-32]\\
~~~~~~~,,       &            &              &                   &                     & 48$-$51        &                     &[33]\\
\noalign{\smallskip}\hline\noalign{\smallskip}
GS 1124-684     &  K3/5 V    & {\bf 0.4326058(31)}& {\bf 3.01$\pm$0.15} & {\bf 0.13$\pm$0.04}   &                &            &[34-36]\\
~~~~~~~,,       &            &              &                   &                     & 60$^{+5}_{-6}$ &  5.0$-$7.5          &[36]\\
~~~~~~~,,	&	     &	            &                   &                     & 54$^{+20}_{-15}$& 5.8$^{+4.7}_{-2.0}$ &[37]\\
~~~~~~~,,	&	     &              &                   &                     & 54$\pm$2       &  7.0$\pm$0.6	     &[38]\\
\noalign{\smallskip}\hline\noalign{\smallskip}
GS 2000+250      &  K3/7 V    & 0.3440915(9) & {\bf 5.01$\pm$0.15}  & {\bf 0.042$\pm$0.012}   &                &                     &[39-41]\\
~~~~~~~,,       &            &	            &                   &                     & 65$\pm$9       &  8.5$\pm$1.5        &[42]\\
~~~~~~~,,       &            &	            &                   &                     & 43$-$69        &  4.8$-$14           &[43]\\
~~~~~~~,,       &            & {\bf 0.3440873(2)} &           &      & {\bf 58$-$74}        &  {\bf 5.5$-$8.8}          &[44]\\
\noalign{\smallskip}\hline\noalign{\smallskip}
A0620-00        &  K2/7 V    & 0.323014(1)  &  2.91$\pm$0.08    &                     &    $<$50       &  $>$7.3             &[45-46]\\
~~~~~~~,,       &            &              &  2.72$\pm$0.06    & {\bf  0.067$\pm$0.010}  &            &                     &[47]\\
~~~~~~~,,       &            &              &  2.76$\pm$0.04    &  0.060$\pm$0.004    &                &                     &[48]\\
~~~~~~~,,       &     & {\bf 0.32301405(1)}&  {\bf 2.76$\pm$0.01}   &                       &               &                     &[49]\\
~~~~~~~,,       &            &              &                  &                       & 63$-$74       &  4.1$-$5.4          &[50]\\
~~~~~~~,,       &            &              &                  &                       & 31$-$54       & 10$^{+7}_{-5}$      &[51]\\
~~~~~~~,,       &  K3/7 V    &              &                  &                       & 38$-$75       &  3.3$-$13.6         &[52]\\
~~~~~~~,,       &            &              &                  &                       & 41$\pm$3      & 11.0$\pm$1.9        &[53]\\
~~~~~~~,,       &            &              &                  &    & {\bf 51$\pm$0.9}    & {\bf 6.6$\pm$0.3}         &[54]\\
\noalign{\smallskip}\hline\hline\noalign{\smallskip}
XTE J1650-500   & $\approx$K4 V	& {\bf 0.3205(7)} &  {\bf 2.73$\pm$0.56}   &                       & $>$47         &		     &[55]\\
\noalign{\smallskip}\hline\noalign{\smallskip}
GRS 1009-45     &  K7/M0 V   & {\bf 0.285206(14)} &  {\bf 3.17$\pm$0.12}   &                       &               &                     &[56]\\
~~~~~~~,,       &            &              &                  &                       & 37$-$80       &                     &[57]\\
\noalign{\smallskip}\hline\noalign{\smallskip}
XTE J1859+226   & $\approx$K5 V & {\bf 0.274(2)}  &  {\bf 4.5$\pm$0.6}    &                       & $<$70         & $>$ 5.42            &[58-59]\\
\noalign{\smallskip}\hline\noalign{\smallskip}
GRO J0422+32    &  M0/4 V    & 0.21159(57)  &  1.21$\pm$0.06   & 0.12$^{+0.08}_{-0.07}$&               &                     &[60-63]\\
~~~~~~~,,       &  M4/5 V    & {\bf 0.2121600(2)} &  {\bf 1.19$\pm$0.02}  & {\bf 0.11$^{+0.05}_{-0.02}$}&  $<$45        &  $>$2.2             &[64]\\
~~~~~~~,,	&  M0/4 V    &              &                  &                       & 35$-$49       &  $\sim$2.5$-$5.0    &[61]\\
~~~~~~~,,	&            &              &                  &                       & $\le$45       &                     &[65]\\
~~~~~~~,,       &            &	            &                  &                       & 10$-$26       &  $\gtrsim$15        &[66]\\
~~~~~~~,,       &            &	            &                  &                       &  45$\pm$2     &  3.97$\pm$0.95      &[67]\\
~~~~~~~,,       &            &              &                  &  $<0.04$              &  $<$30        &  $\gtrsim$10.4      &[68]\\
\noalign{\smallskip}\hline\noalign{\smallskip}
XTE J1118+480   & K5/M1 V & {\bf 0.1699339(2)} & {\bf 6.27$\pm$0.04} & {\bf 0.024$\pm$0.009}  &    &               &[69-73]\\
~~~~~~~,,       &            &              &                  &       0.037$\pm$0.007        &    &               &[74]\\
~~~~~~~,,       &            &              &                  &                        &  68$\pm$2    & 8.30$^{+0.28}_{-0.14}$ &[75]\\
~~~~~~~,,       &  K7/M1 V   &              &                  &                        &  {\bf 68$-$79}     & {\bf 6.9$-$8.2}   &[76]\\
\noalign{\smallskip}\hline\noalign{\smallskip}
\end{tabular}
}
{\small REF: [1] \cite{greiner01}; [2] \cite{harlaftis04}; [3] \cite{hurley13}; [4] \cite{steeghs13}; 
[5] \cite{casares92}; [6] \cite{casares94}; [7] \cite{casares96}; [8] \cite{wagner92}; 
[9] \cite{shahbaz94b}; [10] \cite{pavlenko96}; [11] \cite{sanwal96}; [12] \cite{khargharia10}; 
[13] \cite{orosz01}; [14] \cite{bailyn95}; [15] \cite{orosz97}; [16] \cite{vanderhooft98}; 
[17] \cite{greene01}; [18] \cite{beer02}; [19] \cite{shahbaz99}; [20] \cite{buxton99}; 
[21] \cite{shahbaz03a}; [22] \cite{gonzalez08b}; 
[23] \cite{casares04}; [24] \cite{casares09}; [25] \cite{hynes03a}; [26] \cite{munoz08a}; 
[27] \cite{orosz02}; [28] \cite{orosz11a}; [29] \cite{orosz98}; [30] \cite{remillard96}; 
[31] \cite{filippenko97}; [32] \cite{harlaftis97}; [33] \cite{martin95}; 
[34] \cite{remillard92}; [35] \cite{casares97}; [36] \cite{orosz96}; [37] \cite{shahbaz97}; 
[38] \cite{gelino01a}; [39] \cite{casares95b}; [40] \cite{filippenko95b}; 
[41] \cite{harlaftis96}; [42] \cite{callanan96b}; [43] \cite{beekman96}; 
[44] \cite{ioannou04}; [45] \cite{mcclintock86}; [46] \cite{orosz94}; 
[47] \cite{marsh94}; [48] \cite{neilsen08}; [49] \cite{gonzalez10}; 
[50] \cite{haswell93}; [51] \cite{shahbaz94a}; [52] \cite{froning01}; 
[53] \cite{gelino01b}; [54] \cite{cantrell10}; [55] \cite{orosz04}; 
[56] \cite{filippenko99}; [57] \cite{shahbaz96}; [58] \cite{filippenko01}; 
[59] \cite{corral11}; [60] \cite{orosz95}; [61] \cite{casares95a}; 
[62] \cite{filippenko95a}; [63] \cite{harlaftis99}; [64] \cite{webb00}; 
[65] \cite{callanan96a}; [66] \cite{beekman97}; [67] \cite{gelino03}; 
[68] \cite{reynolds07}; [69] \cite{wagner01}; [70] \cite{mcclintock01}; 
[71] \cite{torres04}; [72] \cite{gonzalez08a}; [73] \cite{calvelo09}; 
[74] \cite{orosz01}; [75] \cite{gelino06}; [76] \cite{khargharia13}\\

\noindent
NOTES TO TABLE 1: {\bf GRS 1915+105:} the inclination angle is derived
from the orientation of radio jets \citep{fender99}.  Note that the
radial velocity curve, and thus the BH mass, might be affected by
irradiation because the binary has remained active since its
discovery, although there are currently no signs for such effects.\\
{\bf V404 Cyg:} [10] model the lower envelope of an I-band light curve
to minimize flickering contribution, but do not account for dilution
from non-variable disc light. Optical spectroscopy indicates veiling
is very small at RI wavelengths but the impact in ellipsoidal modeling
should be tested through simultaneous photometric and spectroscopic
observations.  [9] model a K-band light curve assuming no disc
contribution while [11] demonstrate significant flickering is present
in the H-band. [12] fit the H-band light curve of [11] after
correcting for the disc contribution which is, however, estimated from
non-simultaneous NIR spectroscopy. \\
{\bf XTE J1819.3-2525:} limits to the inclination angle are derived
from ellipsoidal  fits to a low quality archival photographic light curve.\\
{\bf GRO J1655-40:} the mass functions reported in [14-15] are likely
biased by irradiation effects.  [19] and [22] quote mass functions in
true quiescence but they disagree at $>3\sigma$ level. The latter is
obtained from an orbital solution with sparse phase coverage,
resulting in a significantly different period and systemic velocity
with respect to previous studies. Therefore, we tentatively favour the
former solution although this issue needs to be investigated further.
[15-18] report $q$ values derived from fits to the ellipsoidal
modulations, the others are obtained from $V \sin i$.  [21] determine 
$q$ by fitting line profiles with synthetic spectra computed using 
NEXTGEN model atmospheres in a Roche geometry, accounting for 
variations in temperature and gravity. We favour this $q$ determination.
[18] fit BVRI light curves from [15] simultaneously, using Kurucz
model atmospheres. The distance and colour excess are not fixed but included 
in the fit as free parameters, and the best (adopted) solution assumes the 
distance from \cite{hjellming95}. We tentatively favour the inclination
reported in [18] although it should be noted that the $q$ value implied
by their ellipsoidal fits is significantly lower than our favoured
value. In any event, there is an excellent agreement between
all inclination values reported by four different groups. This is the
only case where the error on the BH mass is dominated by uncertainties in 
$q$ and the mass function rather than in the inclination. 
We tentatively favour the BH mass reported in [21]
because it is based on both our favoured $q$ and mass
function from [19], which is free from irradiation effects.  The BH
mass quoted in the abstract of [21] is wrong due to a typo
(Shahbaz, private communication).\\
{\bf BW Cir:} the light curves are dominated by strong flickering
which disguise the ellipsoidal modulation. \\
{\bf GX 339-4:} the mass function is based on radial velocities of
Bowen
emission lines emanating on the inner hemisphere of the irradiated donor. \\
{\bf XTE J1550-564:} [28] derive inclination constraints by
simultaneously fitting optical and NIR lightcurves. Conservative BH
masses are quoted after allowing for different values of disc
contribution obtained through non-simultaneous spectroscopy.\\
{\bf 4U 1543-475:} low inclinations are implied by the small amplitude
of some low quality VI-band ellipsoidal light curves.  However, quoted
values should be treated with caution because of possible systematics
derived from model assumptions, namely disc light is neglected and the
early type donor is assumed to be synchronized and filling its Roche lobe.\\
{\bf H1705-250:} the limited spectral resolution in the spectra
reported in [32] only allow one to set an upper limit to $q$.  [30]
derive broad limits to the inclination based on the absence of
eclipses ($i<80^\circ$) and assuming no disc contribution
in a V-band lightcurve ($i>60^\circ$). \\
{\bf GS 1124-684:} [37-38] derive inclinations assuming no disc
contribution to the NIR light curves. The latter authors also quote
uncertainties which seem unrealistically small given potential
systematic effects. While [36] tries to account for disc veiling in
their B$+$V \& I-band lightcurves, large deviations from ellipsoidal
morphology make the derived inclinations suspect. Note also that
this XRT is strongly affected by large amplitude flickering 
\citep{hynes03b,shahbaz10}.\\
{\bf GS 2000+250:} the mass ratio reported in [41] should be treated
with some caution because of the limited spectral resolution of the
spectroscopic observations.  [42-43] neglect any disc contribution in
their ellipsoidal fits to NIR light curves. Inclinations reported in
[44] are tentatively favoured because modeling accounts for a hot-spot
component and the maximum disc light
contribution ($<$32 percent in the R-band) allowed by the absence of X-ray eclipses. \\
{\bf A0620-00:} the $q$ value reported in [47] is slightly preferred
over that in [48] because the former accounts for the non-sphericity of the
Roche lobe, although this correction may not be statistically
significant. In any case both determinations are fully consistent.\\
{\bf GRS 1009-45:} wide limits to the inclination are obtained from the absence
of X-ray eclipses and ellipsoidal fits to an R-band light curve, assuming no disc
contribution. \\
{\bf XTE J1859+226:} light curves are dominated by strong episodic flickering. \\
{\bf GRO J0422+32:} the reported mass ratios should be treated with
some caution because of the limited spectral resolution of the
spectroscopic data. All inclinations reported are suspected because of very large 
flickering amplitude in optical and NIR light curves.\\
{\bf XTE J1118+480:} the $q$ value reported in [73] has been corrected
for orbital smearing of the line due to the motion of the mass donor
during the exposure. Inclination and BH masses given in [76] are
slightly preferred over those in [75] because the latter are based on
multiwavelength ellipsoidal fits to BVRJHK lightcuves which are
non-simultaneous. [76] only fits H-band light curves but the disc
contribution is accounted for via contemporaneous NIR spectroscopy. It
also provides a wider range of inclinations which is seen as more
realistic. A period derivative for the orbital period of
$\dot{P}=$-1.83$\pm$0.66 ms yr$^{-1}$ was derived by \cite{gonzalez12}.\\
}

\section{Dynamical Black Holes in High-Mass X-ray Binaries (HMXBs)}
\label{sec:3}

In addition to the 17 transients listed in Table~\ref{tab:1},
dynamical mass measurements of BHs have also been possible for four
X-ray binaries with high-mass donor stars: Cyg X-1 and the
extragalactic sources LMC X-1, LMC X-3 and M33 X-7. HMXBs are
persistent X-ray sources, with typical X-ray luminosities
$\sim10^{37}$ ergs s$^{-1}$ powered by massive stellar winds or
incipient Roche lobe overflow. Despite being persistent X-ray
sources, irradiation effects are mostly negligible because the
X-ray luminosity is smaller than or comparable to the bolometric 
luminosity of the mass-losing star. Further, the
contribution of the accretion disc to the total optical flux can be
ignored and thus ellipsoidal light curves are not affected by
systematic effects as discussed in the previous section. There are,
however, two important limitations regarding mass determination in
HMXBs. First, the BH mass is very sensitive to uncertainties in the
mass of the optical star. The latter is highly uncertain because donor
stars in HMXBs are typically undermassive for their spectral types due
to secular mass transfer and binary evolution
(\citealt{rappaport83, podsiadlowski2003}; also compare spectral types and donor masses 
implied by Table~\ref{tab:2}). Second, mass transfer in HMXBs is mostly powered
by stellar winds rather than Roche lobe overflow 
and this has a two-fold effect. On the one hand, wind emission can contaminate   
radial velocities from the massive star, specially if these are obtained from 
lower excitation Balmer lines (e.g. \citealt{ninkov87}). On the other hand, if  
the companion star is not filling its Roche lobe then 
one of the previous assumptions breaks down and
$q$ and $i$ can be underestimated if derived from $V \sin i$ and ellipsoidal 
models as before. To circumvent this problem, extra parameters need to be included when
modeling the observations, namely the Roche lobe filling factor and
the degree of synchronization of the companion star
(e.g.~\citealt{gies86, orosz07}). 

\begin{figure}
  \includegraphics[width=0.70\textwidth,angle=0]{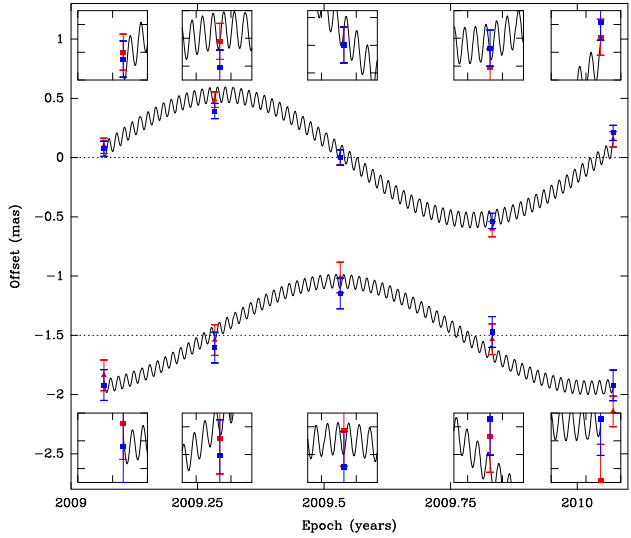}
\caption{Astrometric parallax of the compact radio source in 
Cyg X-1. The long sinusoid indicates the annual parallax after removing the proper
motion of the system. The short period sinusoid reflects the 5.6 day orbit of the 
BH around the center of mass of the binary. From \cite{reid11}.}
\label{fig:8}       
\end{figure}

Despite these caveats, accurate masses can still be obtained if
dynamical constraints are combined with a determination of the radius
of the optical star.  Accurate knowledge of the luminosity and hence
the distance is required which is often difficult. For example, BH
masses reported for Cyg X-1 over the last 3 decades show a large
dispersion, with values between 7-29 M$_{\odot}$, mainly owing to
distance uncertainties. Fortunately, a major advance has been allowed
by the determination of the trigonometric parallax distance of Cyg X-1 with
VLBA. \cite{reid11} have used emission from the compact radio jet to
trace the position of the BH over 1 year. A parallax of
0.539$\pm$0.033 {\it mas} has been derived which in turn implies a
distance of 1.86$\pm$0.12 kpc. The astrometric measurements
are even sensitive to the size of the BH orbit allowing to constrain
its radius to 0.18$\pm$0.09 AU (Fig. \ref{fig:8}). Using the accurate
distance as an extra constraint to the dynamical model, a BH mass of
14.8$\pm$1.0 M$_{\odot}$ is derived, one of the most precise
determinations to date \citep{orosz11b}. The best model solution also
proves that the binary orbit is slightly eccentric ($e\simeq0.02$) and the
companion star rotates faster than the synchronization value at
periastron.

\begin{figure}
  \includegraphics[width=0.50\textwidth,angle=0]{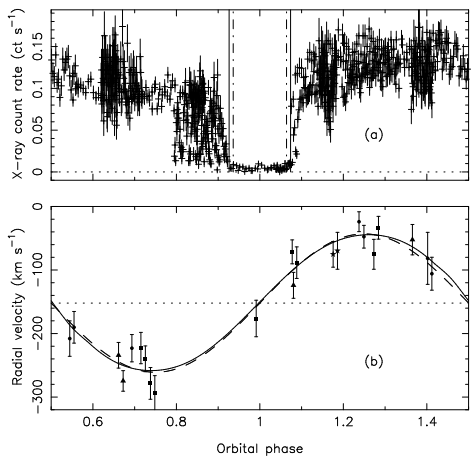}
\caption{ (a) Chandra X-ray light curve of M33 X-7 showing the eclipse of the
compact source by the optical companion. (b) Radial velocity curve of the O7-8III companion star. 
From \cite{orosz07}.}
\label{fig:9}       
\end{figure}

Another remarkable result has been reported for M33 X-7, the first
eclipsing stellar-mass BH.  M33 X-7 is located in the nearby spiral
galaxy M33 and hosts a late O companion which eclipses the X-ray
source every 3.45 days \citep{larson97}.  Ellipsoidal fits to the
optical light curves alone have suggested the presence of a BH
companion \citep{pietsch06} but confirmation through radial velocities
became challenging because of crowding and contamination by nebular
lines. A radial velocity curve of the O7-8III companion was finally
reported in \cite{orosz07}, together with updated ellipsoidal fits
(Fig.~\ref{fig:9}). The duration of the eclipse and the distance to
M33 provides new restrictions which tightly constrain the parameter
space of the dynamical model, in particular the binary inclination,
the companion radius and its filling factor. As in Cyg X-1, the orbit
is found to be slightly eccentric, while the radius of the donor star
extends up to 78 percent of its Roche lobe.  The best fit yields a BH
mass of 15.7$\pm$1.5 M$_{\odot}$, one of the largest accurately
known. The companion star is underluminous for its spectral type and,
with 70.0$\pm$6.9 M$_{\odot}$, it is also one of the most massive
stars known with high accuracy.  M33 X-7 strongly impacts theories of
BH formation and binary evolution since it is very hard to produce
this massive HMXB in such a tight orbit \citep{valsecchi10}.

Table~\ref{tab:2} lists a summary of fundamental parameters in the
four HMXBs with dynamical BH mass measurements. In some cases, early
mass estimates were derived after assuming that the optical star is
synchronized and/or filling its Roche lobe. However, the most accurate
results were subsequently obtained through fitting the ensemble of
light curves, radial velocities and projected rotational velocities
with a full model parametrization. These models account for
non-synchronous rotation, Roche lobe filling fraction and (sometimes)
eccentricity. As mentioned above, a determination of the distance,
which is constrained to better than 6 percent in the four HMXBs, is
key to deliver accurate BH masses. Despite this, a precise BH mass for
LMC X-3 is not yet available because of the large impact of 
X-ray heating in the analysis. Note that we only
quoted BH masses derived using the dynamical mass measurement method
explained above. Mass estimates based on other techniques, such as
fitting model atmospheres \citep{herrero95, caballero-nieves09},
stellar evolutionary models \citep{ziolkowski05} or scaling X-ray
spectral/timing properties \citep{shaposhnikov07}, are not considered
here.  Albeit with low number statistics, Table~\ref{tab:2} shows that
BHs in HMXBs have typically a mass $>$10 M$_{\odot}$ and hence they
tend to be more massive than BHs in XRTs. This seems to hint at an
evolutionary difference, with BHs in HMXBs descending from massive
progenitors which may have experienced comparatively lower stellar
wind mass-loss rates, possibly through case C mass transfer
\citep{wellstein99}. Low metallicity environments, which strongly
influence Wolf-Rayet mass-loss rates and the radius evolution of
progenitor stars, is another important factor in the three
extragalactic HMXBs \citep{crowther10}.  It has also been proposed
that BHs more massive than $\sim$10 M$_{\odot}$ may form through
direct implosion of the core remnant \citep{mirabel03} although it
remains unclear why prompt collapse should have a larger incidence in
HMXBs.

We did not include the mass measurements for the two extra-Galactic
HMXBs NGC~300 X--1 and IC~10 X--1 by \citet{crowther10} and
\citet{silver08}, respectively, as we deem these mass
measurements less secure than those of the four sources we list in
Table~\ref{tab:2}. The main reason for this is that the BH mass is
strongly dependent on the assumed mass of the Wolf-Rayet star. The
latter depends on the assumed stellar luminosity of the Wolf-Rayet
star which in turn can be influenced by contamination from unresolved
stars in the galaxies under study. These two sources are nevertheless
very interesting as they may harbor BHs with the largest masses 
measured so far.

Several authors have examined the mass distribution of stellar 
mass BHs in order to gain insight into BH formation models. 
Bayesian analysis of the observed distribution suggests the presence of 
a mass gap or dearth of compact objects in the 
interval $\sim$2-5 M$_{\odot}$  \citep{ozel10, farr11}. This 
has been interpreted  
in the context of supernova models by a rapidly evolving explosion 
(within ~0.2 s after the core bounce) through a Rayleigh-Taylor instability 
\citep{belczynski12}. On the other hand, the absence of low mass BHs 
has been attributed to a potential observational 
artefact caused by the systematic uncertainties affecting ellipsoidal fits 
and hence inclination measurements (see Section~\ref{sec:2.1}; 
\citealt{kreidberg12}). This has a strong impact on BH formation scenarios 
since confirmation of a mass gap would rule out accretion induced 
collapse while it would support delayed supernova explosion 
models. Future accurate mass measurements of BHs in high-inclination 
systems will allow testing whether the mass gap is caused by systematics 
in the binary inclination or it is a signature of the BH formation 
mechanism. This involves discovering new favourable XRTs and also 
exploiting novel techniques for mass measurements.   

\begin{table}
\caption{Mass measurements of BH in HMXBs. We
  mark in bold face the measurements we consider the most reliable
  currently available in the literature. }
\label{tab:2}       
%
%
\begin{tabular}{lccccccc}
\hline\hline\noalign{\smallskip}
 System & Donor  &  $P_{\rm{}orb}$ &  $f(M)$ & $q$ & i & $M_{\rm x}$   & Ref.\\
 &  Spect. Type & [days] &  [$M_{\odot}$] &    & [deg] &  [$M_{\odot}$] & \\ 
\noalign{\smallskip}\hline\hline\noalign{\smallskip}
Cyg X-1   & $\sim$B0Ib & $\sim$5.607  &   $\sim$0.16      &        &             &  $>$3         & [1-2] \\
~~~~~~~,,     &  09.7 Iab & 5.5995(9)     &   0.182           & $\approx$2.1  & $\approx$30 &  $\approx$14  & [3] \\
~~~~~~~,,     &           & 5.59974(8)    & 0.25$\pm$0.01     & 1.95$-$2.31   & 28$-$38     & 16$\pm$5      & [4-5]\\
~~~~~~~,,     &  & {\bf 5.599829(16)} & {\bf 0.244$\pm$0.006} & {\bf 1.29$\pm$0.15} & {\bf 27.1$\pm$0.8} & {\bf 14.8$\pm$1.0} & [6]\\
\noalign{\smallskip}\hline\noalign{\smallskip}
LMC X-1  & 07 III & 4.2288(6) & 0.14$\pm$0.05 & $\gtrsim$2 & 40$-$63      & $\approx$6   &    [7-8]\\
~~~~~~~,,     & & {\bf 3.90917(5)}  & {\bf 0.149$\pm$0.007} & {\bf 4.91$\pm$0.53} & {\bf 36.4$\pm$1.9} & {\bf 10.9$\pm$1.4} & [9]\\
\noalign{\smallskip}\hline\noalign{\smallskip}
M33 X-7  & 07-8 III  & {\bf 3.453014(20)}  & {\bf 0.46$\pm$0.08} & {\bf 4.47$\pm$0.60} & {\bf 74.6$\pm$1.0} & {\bf 15.7$\pm$1.5} & [10]\\
\noalign{\smallskip}\hline\noalign{\smallskip}
LMC X-3  & B3 V & 1.70491(7) & 2.3$\pm$0.3 &        &   50$-$70    &  7$-$14    & [11]\\
~~~~~~~,,     & B3-5V     & 1.70479(4)    & 2.99$\pm$0.17     &  	      &   50$-$70    &  9.5$-$13.6  & [12]\\
~~~~~~~,,     &   & {\bf 1.7048089(11)} & {\bf 2.77$\pm$0.04 }    &  	      &              &              & [13]\\
\noalign{\smallskip}\hline\noalign{\smallskip}
\end{tabular}
REF: [1] \cite{webster72}; [2] \cite{bolton72a}; [3] \cite{bolton72b}; [4] \cite{gies82}; 
[5] \cite{gies86}; [6] \cite{orosz11b}; [7] \cite{hutchings83}; [8] \cite{hutchings87}; 
[9] \cite{orosz09}; 
[10] \cite{orosz07}; [11] \cite{cowley83}; [12] \cite{val-baker07}; 
[13] \cite{song10} \\

NOTES TO TABLE 2: {\bf Cyg X-1:} [3] uses ellipsoidal light curves from 
\cite{cherepashchuk73} to constrain $q$ and $i$, assuming the star fills the 
Roche lobe. Also assumes a 30 M$_{\odot}$ companion to estimate the BH mass. 
[5] assumes synchronization and fill-out factor in the range 0.9$-$1. 
[6] uses the orbital period determination, radial velocity data and light curves from \cite{brocksopp99}. 
Also adopts a rotational broadening determination from \cite{caballero-nieves09}. \\
{\bf LMC X-1:} [7-8] use a lower limit to $q$ from radial velocities of the NIII $\lambda$4640 emission. 
Synchronization is also assumed to derive a lower limit to the inclination. \\
{\bf M33 X-7:} [10] adopts the orbital period determination from X-ray eclipses \citep{pietsch06}.\\
{\bf LMC X-3:} [11] assumes synchronization to derive a lower limit to the inclination. 
[12] finds evidence of irradiation in the donor star and the BH mass has been corrected for it. \\
\end{table}

\section{Future perspectives}
\label{sec:4}

\begin{figure}
  \includegraphics[width=0.45\textwidth,angle=-90]{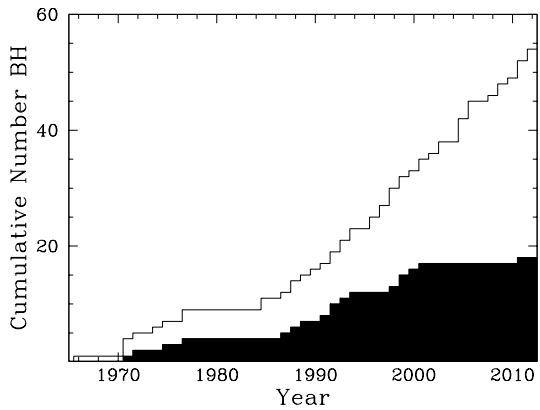}
  \caption{Cumulative distribution of BH XRTs discovered during the
    era of X-ray astronomy. The black histogram indicates XRTs with
    BHs proven dynamically.  Note, however, that the black histogram
    indicates the times when dynamical BHs were first discovered by
    X-ray satellites and not when the actual mass functions were
    measured/reported. The latest dynamical mass measurement BH
    corresponds to Swift J1357.2-0933 although the evidence is
    indirect (see \citealt{corral13}).  Updated after
    \cite{casares10}.  }
\label{fig:10}       
\vspace*{3em}
\end{figure}

The best prospects to enlarge the current sample of stellar-mass BHs
are offered by new discoveries of XRTs. Figure~\ref{fig:10} presents
the cumulative number of BH XRTs discovered in the X-ray astronomy
era, starting with the historic detection of Cen X-2 by a rocket
flight in 1966. A linear increase is apparent since the late 80's,
when X-ray satellites with increased sensitivity and All-Sky-Monitors
became operational. This sets a discovery rate of $\sim$1.7 XRTs yr$^{-1}$.
Extrapolation of the number of XRTs detected to date suggests that
several thousand ``dormant" BHs remain to be discovered
\citep{romani98}.  And this estimate is most likely biased low because
of sample incompleteness and complex selection effects. In particular a
likely population of long period XRTs with very long outburst duty
cycles is often ignored (c.f. \citealt{ritter02}).  In addition, there
is mounting evidence for the existence of a significant number of
intrinsically faint or obscured 
XRTs \citep{degenaar10, corral13}. Incidentally, the latest
population-synthesis models predict $\sim10^3-10^4$ XRTs in the Galaxy
\citep{yungelson06, kiel06}. In either case, the observed sample is 
just the tip of the iceberg of a large "hibernating" population which 
becomes slowly unveiled through outburst episodes 
\citep{vandenheuvel92}. Figure~\ref{fig:10} also shows that only 17 BH 
XRTs have been dynamically confirmed, representing about 30 percent of the 
total number discovered. The remaining ones are virtually lost during decay
to quiescence because they become too faint, even for 10m-class
telescopes.  Therefore, improving on the statistics of BH masses
requires not only a new generation of ELT telescopes to tackle fainter
targets but also new strategies aimed at unveiling a large fraction of
the "hibernating" population of quiescent XRTs. A promissing approach is
presented by the {\it Galactic Bulge Survey} ({\it GBS}), where
potential quiescent BH XRTs are selected in Chandra observations
of regions in the Galactic Bulge 1 degree away from the Galactic plane 
\citep{jonker11, torres13}.

\subsection{Reprocessed Bowen emission}
\label{sec:4.1}

New opportunities for the study of BH XRTs are also opened by the
analysis of emission lines and timing properties of the X-ray reprocessed
radiation. To start with, the discovery of sharp high-excitation
emission lines in active X-ray binaries have demonstrated that dynamical 
information can also be extracted from XRTs while in outburst
\citep{steeghs02}. These lines, first detected in the neutron star binary 
Sco X-1, arise from reprocessing in the 
irradiated companion. The most prominent are the NIII
$\lambda$4634-40 and CIII $\lambda$4647-50 triplets in the core of the
Bowen blend (Fig.~11). In particular, the NIII lines are powered by
fluorescence resonance which requires seed photons of HeII
Ly$\alpha$. Since the velocities of the Bowen lines trace the orbit
of the illuminated hemisphere, a K-correction (which chiefly depends
on the mass ratio and the disc flaring angle) needs to be applied in
order to obtain the true radial velocity curve of the donor star center of
mass \citep{munoz05}. The discovery of sharp Bowen lines during the
2002 outburst of GX 339-4 has allowed the first determination of the
mass function in this Rosetta stone XRT and therefore the dynamical
proof of a BH (\citealt{hynes03a, munoz08a}; see Fig.~11). This work ilustrates 
the power of the technique for systems which otherwise cannot be studied
in quiescence because they become too faint for current
instrumentation.
  
Second, the timing properties of the reprocessed light allows one to
perform Echo Tomography experiments. Echo Tomography exploits time
delays between X-ray and UV/optical variability as a function of
orbital phase to map the illuminated sites in a
binary~\citep{obrien02}. In particular, radiation reprocessed in the
companion star gives rise to a sinusoidal modulation of time lag
versus orbital phase. The shape of the modulation encodes information
on the most fundamental parameters such as the binary inclination,
mass ratio and stellar separation. Therefore, Echo Tomography offers
an alternative route to derive accurate inclinations, which is
critical for measuring BH masses. Unfortunately, attempts to measure
correlated optical/X-ray variability using broad-band filters have
resulted in little evidence for orbital modulation, with time lags
pointing to reprocessing in the outer
disc~\citep{hynes05}. Alternatively, the use of emission line light
curves allows one to amplify the response of the donor's contribution
by suppressing most of the background continuum light, dominated by
the disc. This requires special instrumentation such as {\it ULTRACAM}
\citep{dhillon07}, a high-speed triple-beam CCD camera, equipped with
customized narrow-band filters centered in the Bowen blend and a
nearby continuum. Following this strategy, time lags associated with
the donor star have been finally presented for the neutron star X-ray
binaries Sco X-1 and 4U 1636-536 \citep{munoz07, munoz08b}. In the
latter case, three X-ray/optical bursts were observed at different
orbital phases and their time lags were found to be consistent with
those simulated for a plausible range of binary masses and
inclinations. A careful subtraction of the underlying continuum seems
critical to deliver unbiased inclinations and this requires further
investigation. Echo Tomography has not been attempted on BH XRTs yet
because it is very difficult to coordinate space facilities with
adequate ground-based instruments, such as {\it ULTRACAM}, in ToO
mode. The most efficient use of the technique would require
implementing a fast read-out optical/UV camera onboard of an X-ray
satellite. The camera should be provided with frame transfer EM3CCDs
and a dual-channel, optimized for the Bowen lines (either in the
optical or the higher energy OIII $\lambda$3133/$\lambda$3444
transitions) and an adjacent continuum. Such an instrument would
guarantee both simultaneity and an optimal continuum subtraction.

\begin{figure}[ht]
\begin{center}
\begin{picture}(250,190)(50,30)
\put(0,0){\includegraphics{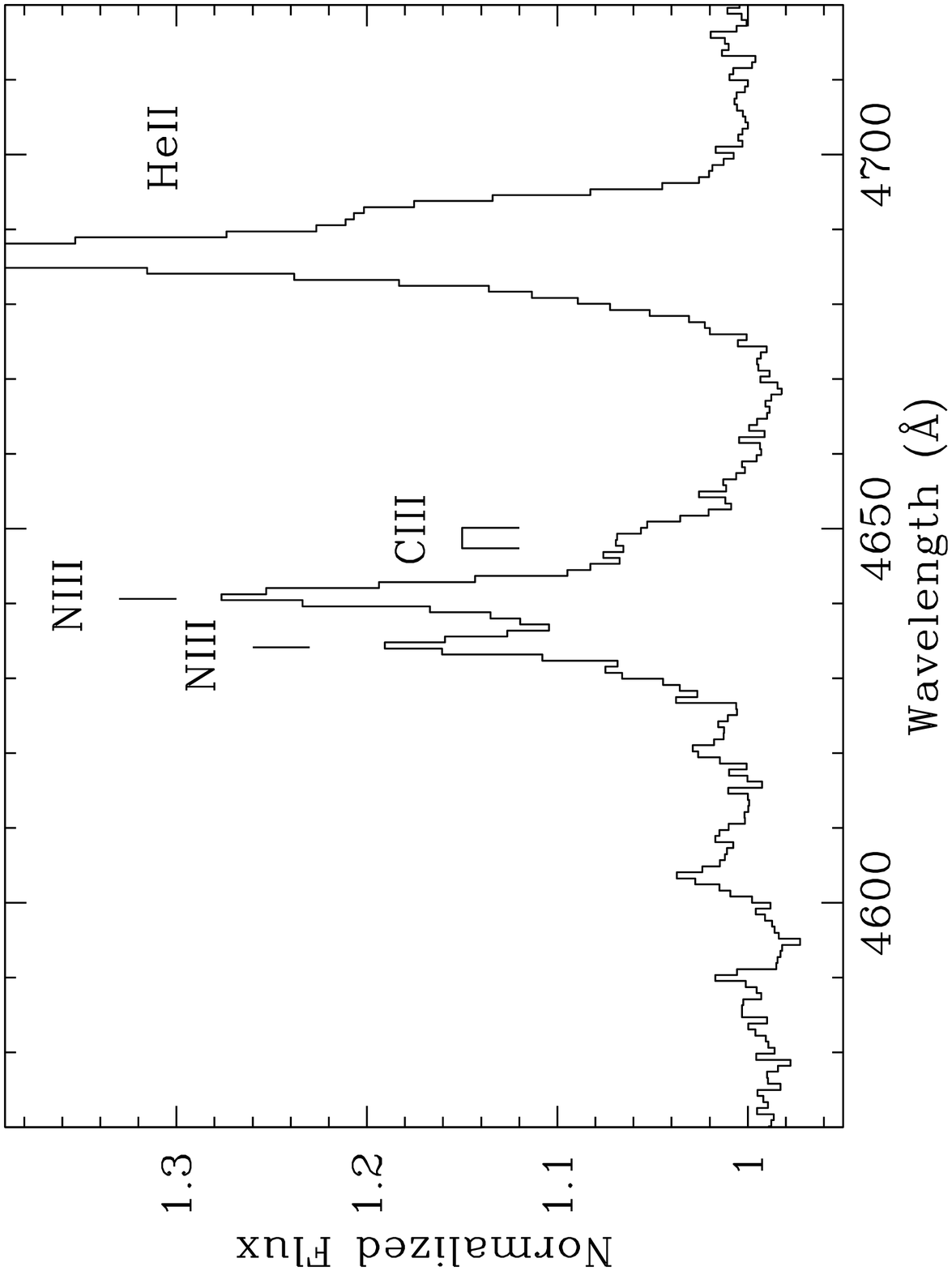}}
\put(0,0){\includegraphics{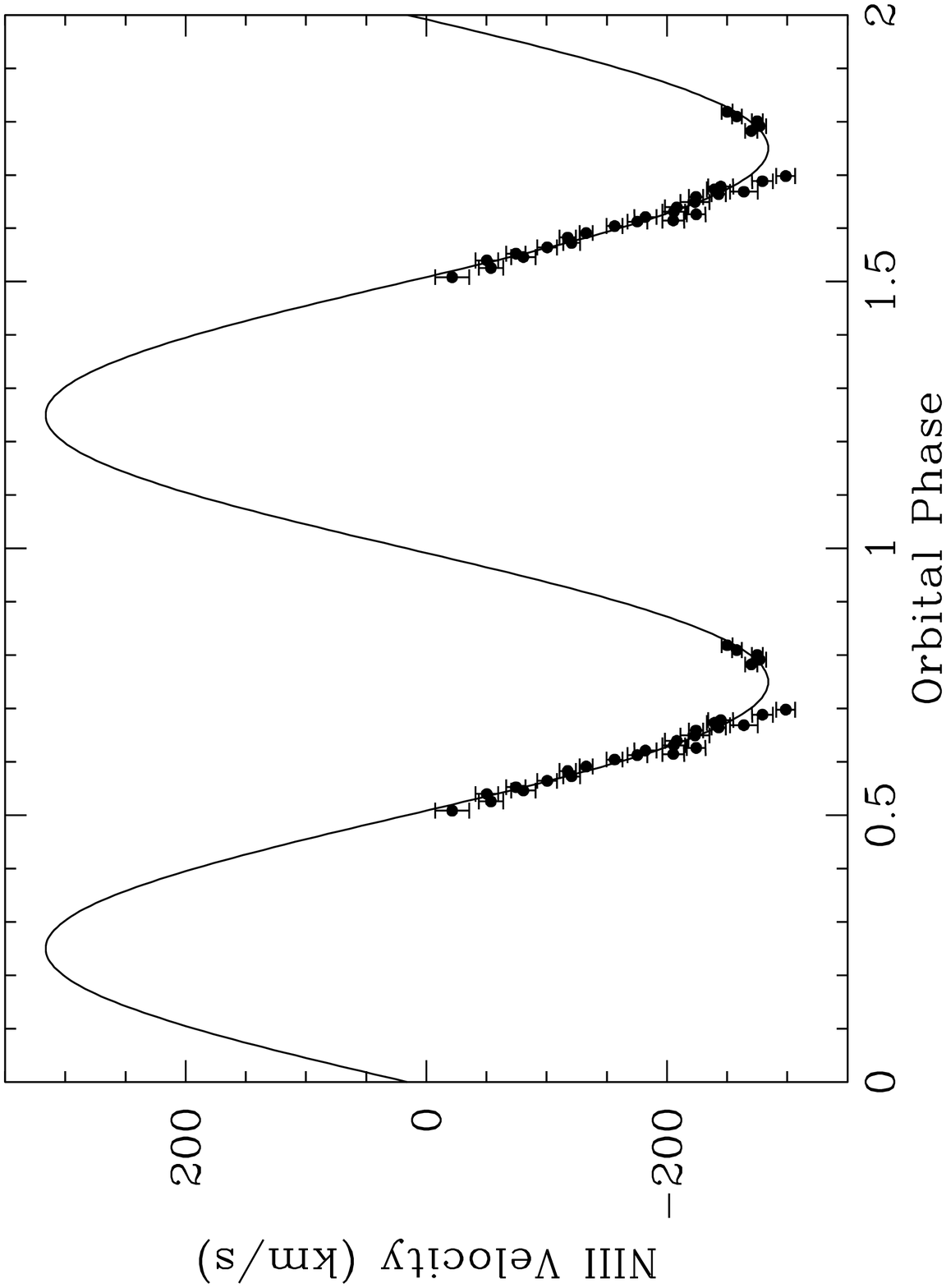}}
\noindent
\end{picture}
\end{center}
{\caption{Detecting companion stars in persistent X-ray binaries and XRTs in outburst. 
Left: high excitation emission lines from the irradiated donor in GX~339-4. 
Right: radial velocity motion of the Bowen CIII/NIII emission lines in 
GX~339-4 as a function of time. Adapted from \cite{hynes03a}.\newline}}
\label{fig:11}       
\vspace*{3em}
\end{figure}
 
\subsection{Optical/NIR Interferometry}
\label{sec:4.2}

A different avenue to obtain accurate inclinations relies on measuring the astrometric orbits 
of the companion stars to the BHs. Table ~\ref{tab:3} lists the angular size ($a/d$) of the 
companion's orbits for the best candidates, with values ranging between 20-60 $\mu$as. 
Resolving the projected orbits thus requires astrometry at micro-arcsec level which 
is technically challenging. Fortunatelly, prospects are bright thanks to new developments on 
optical/NIR interferometry in ground-based and space instrumentation.  
Three main facilities will be capable of measuring the orbits of some of the 
targets listed in Table~\ref{tab:3}: {\it GRAVITY}, {\it GAIA} and the proposed 
{\it Space Interferometry Mission SIM Lite} \citep{unwin08}. 
{\it GRAVITY}, a second generation instrument on VLTI, is scheduled to see first 
light in 2014 \citep{kendrew12}. With a goal of 10 $\mu$as precision astrometry for 
targets brighter than K=15 it can potentially resolve the companion's orbit in GRS 
1915+105. V404 Cyg and Cyg X-1 are also within reach, although pushing the
observations to high airmasses.  
On the other hand, {\it SIM Lite}
is designed to deliver micro-arcsecond astrometry in the optical band down to $V=19$. 
While Cyg X-1 can be easily tackled, V404 Cyg is at the limit because of its optical 
faintness. Finally, {\it GAIA} will allow to measure the orbital 
motion of the companion star in Cyg~X-1. 
The first direct measurements of binary orbits will clearly represent a major step 
forward in the knowledge of the fundamental parameters of these systems.   
Not only BH masses could be determined to better than 10 percent accuracy, 
but also precise inclinations for a sample of BHs will allow to
quantify the impact of systematic errors in ellipsoidal light curve
modeling.  In addition, interferometry will deliver accurate distances
and proper motions, key quantities to determine the BH natal kick and
thus constrain the formation and evolution history of accreting BH
X-ray binaries (e.g. \citealt{wong12}).  Currently, parallax distances
are only available for Cyg X-1 \citep{lestrade99, reid11} and V404 Cyg
\citep{miller-jones09} through radio Very Long Baseline Interferometry
observations. The advent of the optical/NIR interferometry era will
allow one to extend these measurements to any (sufficiently bright) BH
binary in the Galaxy.

\begin{table}
\caption{Best BH candidates for astrometric orbit determination}
\label{tab:3}       
%
%
\begin{tabular}{lccccc}
\hline\hline\noalign{\smallskip}
 System &   $P_{\rm orb}$ &  d     & $a/d$ & V mag   & K mag       \\
        &      [days]      &  [kpc] & [$\mu$as] &         &             \\ 
\noalign{\smallskip}\hline\hline\noalign{\smallskip}
V404 Cyg             & 6.5  &  2.4 &  61 & 18.4  & 12.4 \\
Cyg X-1              & 5.6  &  1.9 &  45 & 8.9   & 6.5  \\   
GRS 1915+105         & 33.5 & 11.5 &  40 & $>$25 & 13.0 \\
A0620-00             & 0.33 &  1.1 &  20 & 18.3  & 14.5 \\
GRO J1655-40         & 2.6  &  3.2 &  18 & 17.2  & 13.2 \\
\noalign{\smallskip}\hline\noalign{\smallskip}
\end{tabular}
\end{table}

\section{Intermediate-mass black holes?}
\label{sec:6}
As one has seen above, stellar black holes with masses up to $\sim
16M_{\odot}$ (see Table~\ref{tab:1} and \ref{tab:2}) have been
found. Theoretically, the mass of a stellar mass black hole depends on
the initial mass of the progenitor, how much mass is lost during the
progenitor's evolution and on the supernova explosion mechanism
(\citealt{2010ApJ...714.1217B}; \citealt{fryer12}). Mass is lost
through stellar winds, the amount of mass lost strongly depends on the
metallicity of the star. For a low--metallicity star ($\sim0.01$ of
the solar metallicity) it is possible to leave a black hole of $
\approxlt 100\ M_{\odot}$ (\citealt{2010ApJ...714.1217B}).

Thus, these models do allow for more massive stellar--mass black holes
than have been found so far in our Galaxy. In current stellar
evolution models the mass distribution of black hole remnants from
massive stars peaks around $10\ M_{\odot}$, with a tail up to the said
$\approxlt 100\ M_{\odot}$.  In very massive stars ($\gtrsim\
130M_{\odot}$) the production of free electrons and positrons, due to
the increased gamma ray radiation, reduces the thermal pressure inside the
core. This eventually leads to a runaway thermonuclear explosion that
completely disrupts the star without leaving a black hole, causing the
upper limit for a stellar black hole of $\sim100\ M_{\odot}$.

Supermassive black holes (SMBHs) in AGN with masses of
$>10^5M_{\odot}$ (\citealt{2007ApJ...670...92G}) have been identified.
However, black holes with masses of several hundred to a few thousand
solar masses remain elusive.  Such intermediate-mass black holes
(IMBHs) may be remnants from the first population of zero-metallicity
massive stars. These Population III stars could have had masses above
the pair--instability limit of around 130 M$_\odot$ and they may have
collapsed into IMBHs (\citealt{2001ApJ...551L..27M}). It has also been
suggested that IMBHs may form in the centers of dense stellar clusters
via the merging of stellar-mass black holes (e.g.,
\citealt{2002MNRAS.330..232C}), or from the collapse of merged
supermassive stars in very dense star clusters (e.g.,
\citealt{2002ApJ...576..899P}).  Primordial formation of a population
of IMBHs is not ruled out (\citealt{2003ApJ...594L..71A}), although
they do not form a significant fraction of the dark matter
(\citealt{2008ApJ...680..829R}).

IMBHs could allow for the assembly of supermassive black holes early
in the Universe (e.g.~\citealt{2010Natur.466.1049V};
\citealt{2012Sci...337..544V}). IMBHs may contain a similar or even a
larger fraction of the baryonic matter than SMBH and they are potential
sources of gravitational waves when they spiral into SMBHs
(\citealt{2001ApJ...551L..27M}). Those IMBHs that were produced early
in the life of the Universe but that did not evolve into SMBHs should
still be wandering in the (outer) halos of galaxies, such as our Milky
way (\citealt{2012MNRAS.421.2737O}; \citealt{2013arXiv1303.3929R}).

So, there is significant interest in finding these IMBHs and
observationally determine their properties. A comprehensive review of
constraints on the existence of IMBHs can be found in
\cite{2003AIPC..686..115V}. The main problem in finding definitive
proof for the existence of IMBHs is that the radius of their sphere of
influence is small ($r_{infl} = \frac{GM_{BH}}{v^2}$). The meaning of
$v$ depends on the case under consideration: v can be the velocity of a
recoiling BH with respect to that of surrounding stars, or for BHs in the centre
of a stellar association/cluster it is the velocity dispertion of the
stars). Below, we will revisit the possible observational evidence
for the existence of IMBHs.

\subsection{IMBHs in globular clusters?}

Given that several formation mechanisms for IMBHs seem to require
dense stellar environments
(\citealt{2002MNRAS.330..232C}; \citealt{2002ApJ...576..899P}), much
effort has gone into investigating whether IMBHs are present in the
cores of globular clusters. 

\subsubsection{Photometric and kinematic evidence}
The number of stars as a function of distance (R) to the centre of a
globular cluster has been used to investigate whether a central BH is
present or not: this function has the form of N(R)$\propto R^{-0.75}$
(\citealt{1976ApJ...209..214B};
\citealt{1977ApJ...216..883B}). However, a globular cluster that has
gone through core collapse (\citealt{1986ApJ...305L..61D}) will show a
similar distribution of the number of stars as a function of radius
from the centre of the globular cluster (e.g.~ the case of M~15 see
\citealt{1992ApJ...392...86G}).

Instead, using both spectroscopic as well as photometric data and
comparing that to a model of stellar orbits in the globular cluster
potential one can derive whether a central IMBH is present or not.  The
observed surface brightness profile together with an assumed
mass-to-light ratio and a central IMBH of a given mass provides a
model for the kinematics of the stars. This model is compared with the
actual kinematics data such as the radial velocity profile and the
velocity dispersion as a function of the distance to the cluster
centre. The data is fit for the mass-to-light ratio and the mass of
the IMBH in the cluster core (which can be zero). IMBHs, if present in
the centre of a globular cluster, will increase the velocity
dispersion in the core. This method has been used to argue for the
presence of IMBHs in a sample of globular clusters (most notably G1
near M~31; \citealt{2002ApJ...578L..41G};
\citealt{2005ApJ...634.1093G}). The presence of compact objects (white
dwarfs, neutron stars and stellar-mass BHs that sink to the centre of
the cluster due to dynamical friction) does not significantly alter
the conclusion about the presence of an IMBH in the centre of the
cluster (\citealt{2005ApJ...634.1093G}). However, conflicting reports
on modeling the observable data exist: \cite{2003ApJ...589L..25B}
obtain good fits to the spectroscopic and the photometric
data using N-body calculations without the need of an IMBH (although the
work of \citealt{2005ApJ...634.1093G} challanges
this). \cite{2013A&A...555A..26L} compile the globular clusters for
which evidence exists that they harbor IMBHs and they further show
that these IMBHs do not follow the M$- \sigma$ correlation as found to SMBHs
(\citealt{2000ApJ...539L...9F}).

\subsubsection{Radio -- X-ray correlation}

For G1, besides the dynamical evidence for the presence of an IMBH,
\cite{2006ApJ...644L..45P} showed that there was X-ray emission of a
source associated with the globular cluster, however, the spatial
scale of their \xmm\, observation did not allow the authors to claim
that the source resides in the core of G1. The source luminosity is
both consistent with that expected for an IMBH accreting at a low
(Bondi-Hoyle) rate as well as with the source being a low-mass X-ray
binary (LMXB) accreting via Roche lobe overflow from a companion star.
\cite{2010MNRAS.407L..84K} improved the X-ray position of the source
using a \chan\, observation. The source position is consistent with
being equal to the centre of the globular cluster.

\cite{2007ApJ...661L.151U} provided evidence for the detection of
radio emission from the same location as the X-ray emission. Given
that the relation between BH mass, X-ray and radio luminosity appears
to follow a “fundamental plane,” in which the ratio of radio to X-ray
luminosity increases as the ∼0.8 power of the BH mass, an IMBH is more
radio loud at a given X-ray luminosity than a stellar-mass BH
(\citealt{2003MNRAS.345.1057M}; \citealt{2004A&A...414..895F}). The
result of \cite{2007ApJ...661L.151U} is consistent with an IMBH
scenario but not with an LMXB scenario. However, the radio and X-ray
observations were not simultaneous and the significance of the radio
detection warranted further investigation. \cite{2012ApJ...755L...1M}
conducted the experiment where simultaneous X-ray (\chan) and radio
(VLA) observations of G1 were obtained. Whereas the X-ray luminosity
was consistent with that found before, these authors find no evidence
for radio emission at the position of the X-ray source down to a limit
of 4.7 $\mu$Jy per beam.  Using the fundamental plane of BH activity
this yields an upper limit on the mass of the IMBH in G1 of
$<9.7\times 10^3$M$_\odot$. Note that this upper limit is only
marginally consistent with the mass of the IMBH derived dynamically
(\citealt{2005ApJ...634.1093G}).  \citet{2012ApJ...750L..27S} used
deep VLA observations of the globular clusters M~15, M~19 and M~22 to
search for radio emission associated with low-level accretion onto an
IMBH. No radio sources were detected, putting additional constraints
on any IMBH in these globular clusters. Overall, it is probably fair
to state that IMBHs have not yet been detected beyond doubt in
globular clusters although the evidence for their existence seems to 
be growing.

Incidently, in the process of searching for IMBHs in globular
clusters, \citet{2012ApJ...750L..27S} {\it did} find evidence for
radio sources in globular clusters albeit not in their cores. The
properties of some of these radio sources (flat spectrum radio
emission and the limits on/detection of X-ray emission) did provide
evidence for the presence of stellar-mass BHs in globular clusters
(\citealt{2012Natur.490...71S}; \citealt{2013arXiv1306.6624C}).

\subsection{Ultraluminous X--ray sources}

Ultraluminous X--ray sources (ULXs) are off-nuclear X-ray point
sources in nearby galaxies with X--ray luminosities, L${_X}\approxgt
1\times 10^{39}$ -- $10^{42}$ erg s$^{-1}$
(e.g.~\citealt{1999ApJ...519...89C}). Their X--ray luminosities are
suggestive of IMBHs if they radiate isotropically at sub--Eddington
levels. Hence, ULXs could harbor IMBHs. Alternatively, the radiation
in ULXs is not emitted isotropically or the Eddington limit is
breached. So called slim disc models could potentially allow for
the latter. Below we discuss these possible explanations for the high
luminosity in ULXs in more detail.

\subsubsection{Beaming and super-Eddington accretion}
Relativistic beaming has been proposed as the cause for the high
apparent luminosity of ULXs (\citealt{2002A&A...382L..13K}). This
model predicts that for every high--luminosity source, there should be
a larger number of lower luminosity sources; around 30 sources of $
L_x\sim 10^{39}$ erg s$^{-1}$ for every source of $L_x\sim 10^{40}$
erg s$^{-1}$. However, approximately 5-10 sources at $10^{39}$ erg
s$^{-1}$ are found for each source with luminosity $10^{40}$ erg
s$^{-1}$ (\citealt{2011MNRAS.416.1844W}). As this scenario requires a
large number of jet sources beamed in other directions, for which
there is no observational evidence it cannot explain the high
luminosity for the majority of sources.

Geometrical beaming, where the emission is non--isotropic, provides,
in combination with super--Eddington accretion, a viable explanation
of the high ULX luminosity. The theory of super--Eddington black hole
accretion was developed in the 1980s with the slim disk model of
\citet{1980ApJ...242..772A}. Here, narrow funnels along the rotation
axis of the accretion disk, in part caused by massive
radiation--driven winds from the accretion disk, collimate radiation
into beams, resulting in an apparent high luminosity as a combination
of collimation and super--Eddington accretion rates. Recent
simulations support this idea and even with a moderately
super--critical mass supply an apparent luminosity $\approx 20\
L_{Edd}$ can be reached (\citealt{2011ApJ...736....2O}).  It has also
been suggested that strong density inhomogeneities in the accretion
disk could cause the escaping flux to exceed the Eddington limit by a
factor of $\sim 10-100$ (\citealt{2002ApJ...568L..97B}).

In these models ordinary stellar mass black holes can reach the
luminosity observed for most ULXs. However, these scenarios do mean
that the Eddington limit is violated and, so far, the Eddington limit
works well for almost all known Galactic black holes and AGN
(\citealt{2010MNRAS.408.1714R}). Some Galactic black holes do seem to
approach or even breach the Eddington limit, but if so, they do so for
only a brief period in time. The conclusion on this issue for Galactic
black hole binaries is hampered by the uncertain distance for many
sources (\citealt{2004MNRAS.354..355J}). Of course, perhaps the
limited inflow rate is causing the AGNs and most X--ray binaries to
remain below the Eddington rate instead of there being a physical
barrier at the Eddington limit (\citealt{2005MNRAS.356..401R}).

The ULXs with the largest luminosities cannot easily be explained by
the geometrical beaming model with super--Eddington accretion rates
onto stellar mass black holes unless the mass of these stellar mass
black holes is larger than those found in our own Galaxy. The higher
the luminosity, the less likely it is that they can be explained as
the high--luminosity end of the X--ray binary stellar-mass BH distribution.

A strong case for an IMBH is provided by the
variable, very bright ULX ESO~243--49 X--1
(\citealt{2009Natur.460...73F}). Its X--ray luminosity is too high for
a stellar--mass black hole even in the presence of some beaming. Given
the evidence for the detection of a redshifted H$\alpha$ emission line
in the optical counterpart to ESO~243--49 X--1
(\citealt{2010ApJ...721L.102W}), the uncertainty on its distance is
reduced with respect to other bright ULXs. Another case for an IMBH is
M82~X41.4+60, whereas it does not reach peak luminosities as high as
ESO~243--49 X--1, the maximum luminosity is still uncomfortably high
for a stellar--mass black hole (\citealt{2003ApJ...586L..61S}).

Are ULXs really IMBHs or stellar--mass black holes under peculiar
accretion conditions? The answer to this question relies on dynamical
mass measurements for the black holes in these systems similar to
those available for stellar-mass BHs (see Sections 2 and 3). Given the
faintness of the optical counterparts (typically V $\geq$ 22 mag; see
for instance \citealt{2004ApJ...602..249L} and
\citealt{2008MNRAS.387...73R}), radial velocity studies of ULXs have
concentrated on strong emission lines in the optical
spectrum. However, these attempts to provide ULX dynamical masses have
not met with success because the emission lines are originating in the
accretion disc or a wind, and not in the donor star itself
(cf.~\citealt{2012ApJ...745...89L};
\citealt{2011AN....332..398R}). One potential exception has been
mentioned by \citet{2009ApJ...704.1628L} who interpret several
emission lines as coming from a Wolf Rayet mass donor star to the ULX
X--1 in M~101, although no follow--up work has been presented so far.

Searches for photospheric lines have so far concentrated on the blue
part of the spectrum as many work on the hypothesis that the donor
stars are blue supergiants. This is based on the fact that some ULXs
are near a young star cluster and by the blue colours of ULX optical
counterparts. However, the blue colours are also consistent with
emission from accretion disks as recently has been confirmed by
\citet{2012ApJ...750..152S} for a ULX (see also
\citealt{2012arXiv1208.4502J} and \citealt{2012ApJ...745..123G} for
discussions). In fact, it could well be that a significant fraction of
the donor stars are red supergiants (\citealt{2005MNRAS.362...79C};
\citealt{2007MNRAS.376.1407C}) which are intrinsically bright in the
infrared, (M${_V} \sim -6$, V--H$\sim 4$, H--K$\sim 0$) in contrast
with the blue supergiants (M${_V} \sim -6$, V--H$\sim 0$, H--K$\sim
0$). Therefore, some ULX systems may resemble the bright Galactic
X-ray binary GRS~1915$+$105 that has a red giant donor star
(\citealt{2001A&A...373L..37G}) and thereby radial velocity
measurements from infrared photospheric lines will be possible.

\begin{figure}[ht]
\begin{center}
\begin{picture}(250,190)(50,30)
\put(0,0){\includegraphics{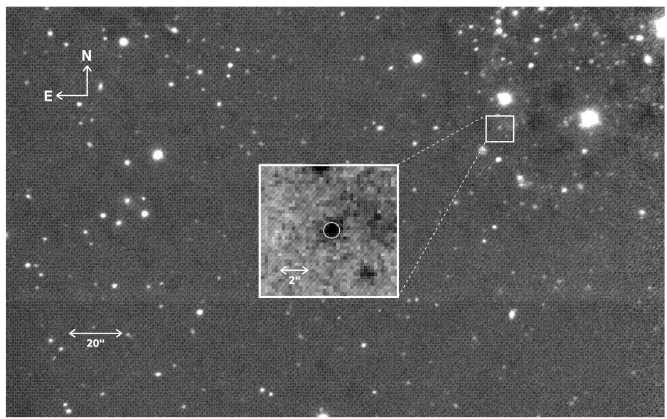}}
\noindent
\end{picture}
\end{center}
{\caption{William Herschel Telescope observations in the Ks-band of
    the ULX Holmberg II X-1. The small circle indicates the \chan\,
    position of the ULX. The inset shows the unique Ks-band
    counterpart to the ULX (from Heida et al. in prep.).\newline}}
\label{fig:12}       
\vspace*{3em}
\end{figure}

From a near-infrared survey of ULX sources within 10 Mpc, Heida et al. 
(in prep.) found evidence (e.g.~see Figure 12) that about 10-15 percent 
of ULXs have a bright near-infrared
counterpart which are consistent with a red-supergiant companion
star. However, future near-infrared spectroscopy of these targets
should be done to investigate the nature of these near-infrared counterparts.

\begin{acknowledgements}
We thanks Cristina Zurita, Andrew Cantrell and Rob Hynes for 
providing us with figures 5, 6 and 11, respectively. JC acknowledges 
support by the Spanish Ministerio de Econom\'\i{}a y Competitividad 
(MINECO) under grant AYA2010--18080.
\end{acknowledgements}


\begin{thebibliography}{}

\bibitem[\protect\citeauthoryear{Abramowicz et
    al.}{1980}]{1980ApJ...242..772A}
M.~A. Abramowicz, M. Calvani, L. Nobili, \mbox{ApJ} \textbf{242}, 772 (1980).
\bibitem[\protect\citeauthoryear{Afshordi, McDonald and
    Spergel}{2003}]{2003ApJ...594L..71A}
N. Afshordi, P. McDonald, D.~N. Spergel, \mbox{ApJ} \textbf{594}, L71
(2003).
\bibitem[\protect\citeauthoryear{Armas Padilla et al.}{2013a}]{armas13a}
M. Armas Padilla, P. N. Degenaar, D.M. Russell, R. Wijnands, \mbox{MNRAS} \textbf{428}, 3083
(2013a).
\bibitem[\protect\citeauthoryear{Armas Padilla et al.}{2013b}]{armas13b}
M. Armas Padilla, R. Wijnands, D. Altamirano, M. M\'endez, J.M. Miller, N. Degenaar, 
\mbox{MNRAS}, arXiv:1308.4326 (2013b, submitted).
\bibitem[\protect\citeauthoryear{{Bahcall} and
  {Wolf}}{1976}]{1976ApJ...209..214B}
\begin{barticle}
\bauthor{\binits{J.N.} \bsnm{{Bahcall}}},
\bauthor{\binits{R.A.} \bsnm{{Wolf}}},
\bjtitle{\apj}
\bvolume{209},
\bfpage{214}--\blpage{232}
(\byear{1976}).
\end{barticle}
\endbibitem

\bibitem[\protect\citeauthoryear{{Bahcall} and
  {Wolf}}{1977}]{1977ApJ...216..883B}
\begin{barticle}
\bauthor{\binits{J.N.} \bsnm{{Bahcall}}},
\bauthor{\binits{R.A.} \bsnm{{Wolf}}},
\bjtitle{\apj}
\bvolume{216},
\bfpage{883}--\blpage{907}
(\byear{1977}).
\end{barticle}
\endbibitem

\bibitem[\protect\citeauthoryear{Bailyn et al.}{1995}]{bailyn95}
C.D. Bailyn, J.A. Orosz, J.E. McClintock, R.A. Remillard,
\mbox{Nature} \textbf{378}, 157 (1995).

\bibitem[\protect\citeauthoryear{{Baumgardt}
  et~al.}{2003}]{2003ApJ...589L..25B}
\begin{barticle}
\bauthor{\binits{H.} \bsnm{{Baumgardt}}},
\bauthor{\binits{J.} \bsnm{{Makino}}},
\bauthor{\binits{P.} \bsnm{{Hut}}},
\bauthor{\binits{S.} \bsnm{{McMillan}}},
\bauthor{\binits{S.} \bsnm{{Portegies Zwart}}},
\bjtitle{\apjl}
\bvolume{589},
\bfpage{25}--\blpage{28}
(\byear{2003}).
\end{barticle}
\endbibitem


\bibitem[\protect\citeauthoryear{Beekman et al.}{1996}]{beekman96}
G. Beekman, T. Shahbaz, T. Naylor, P.A. Charles, \mbox{MNRAS} \textbf{281}, L1 (1996).
\bibitem[\protect\citeauthoryear{Beekman et al.}{1997}]{beekman97}
G. Beekman, T. Shahbaz, T. Naylor, P.A. Charles, R.M. Wagner, P. Martini, \mbox{MNRAS} \textbf{290}, 303 (1997).
\bibitem[\protect\citeauthoryear{Beer \& Podsiadlowsky}{2002}]{beer02}
M.E. Beer, P. Podsiadlowsky, \mbox{MNRAS} \textbf{331}, 351 (2002).

\bibitem[\protect\citeauthoryear{{Begelman}}{2002}]{2002ApJ...568L..97B}
\begin{barticle}
\bauthor{\binits{M.C.} \bsnm{{Begelman}}},
\bjtitle{\apjl}
\bvolume{568},
\bfpage{97}--\blpage{100}
(\byear{2002}).
\end{barticle}
\endbibitem


\bibitem[\protect\citeauthoryear{Belczynski et~al.}{2010}]{2010ApJ...714.1217B}
K. Belczynski, T. Bulik, C.~L. Fryer, A. Ruiter, F. Valsecchi,
J.~S. Vink, J.~R. Hurley, \mbox{ApJ} \textbf{714}, 1217 (2010).
\bibitem[\protect\citeauthoryear{Belczynski et al.}{2012}]{belczynski12}
K. Belczynski, G. Wiktorowicz, C.L. Fryer, D.E. Holz, V. Kalogera, \mbox{ApJ} \textbf{757}, 91 (2012).
\bibitem[\protect\citeauthoryear{Belloni et al.}{2011}]{belloni11}
T.M. Belloni, S.E. Motta, T. Mu\~noz-Darias, \mbox{Bull. Astr. Soc. India}
\textbf{39}, 409 (2011).
\bibitem[\protect\citeauthoryear{Bolton}{1972a}]{bolton72a}
C.T. Bolton, \mbox{Nature} \textbf{235}, 271 (1972a).
\bibitem[\protect\citeauthoryear{Bolton}{1972b}]{bolton72b}
C.T. Bolton, \mbox{Nature} \textbf{240}, 124 (1972a).
\bibitem[\protect\citeauthoryear{Brocksopp et al.}{1999}]{brocksopp99}
C. Brocksopp, A.E. Tarasov, V.M. Lyuty, P. Roche, \mbox{A\&A} \textbf{343}, 861 (1999).
\bibitem[\protect\citeauthoryear{Buxton \& Vennes}{1999}]{buxton99}
M. Buxton, S. Vennes, \mbox{PASP} \textbf{18}, 91 (1999).
\bibitem[\protect\citeauthoryear{Caballero-Nieves et al.}{2009}]{caballero-nieves09}
S.M. Caballero-Nieves et al., \mbox{ApJ} \textbf{701}, 1895 (2009).
\bibitem[\protect\citeauthoryear{Callanan et al.}{1996a}]{callanan96a}
P.J. Callanan, M.R. Garcia, J.E. McClintock, P. Zhao, R. Remillard, F. Haberl, \mbox{ApJ} \textbf{461}, 351 (1996a).
\bibitem[\protect\citeauthoryear{Callanan et al.}{1996b}]{callanan96b}
P.J. Callanan, M.R. Garcia, A.V. Filippenko, I. McLean, H. Teplitz, \mbox{ApJ} \textbf{470}, L57 (1996b).
\bibitem[\protect\citeauthoryear{Calvelo et al.}{2009}]{calvelo09}
D.E. Calvelo, S.D. Vrtilek, D. Steeghs, M.A.P. Torres, J. Neilsen, A.V. Filippenko, J.I. Gonz\'alez Hern\'andez, 
\mbox{MNRAS} \textbf{399}, 539 (2009).
\bibitem[\protect\citeauthoryear{Cantrell et al.}{2008}]{cantrell08}
A.G. Cantrell, C,D, Bailyn, J.E. McClintock, J.A. Orosz, \mbox{ApJ} \textbf{673}, L159 (2008).
\bibitem[\protect\citeauthoryear{Cantrell et al.}{2010}]{cantrell10}
A.G. Cantrell et al., \mbox{ApJ} \textbf{710}, 1127 (2010).
\bibitem[\protect\citeauthoryear{Casares}{1996}]{casares96}
J. Casares, \mbox{IAU Colloq.} \textbf{158}, 395 (1996).
\bibitem[\protect\citeauthoryear{Casares}{2007}]{casares07}
J. Casares, \mbox{IAU Symp.} \textbf{238}, p.3 (2007).
\bibitem[\protect\citeauthoryear{Casares}{2010}]{casares10}
J. Casares, \mbox{Ap\&SS Proc.}, Springer-Verlag Berlin Heidelberg, p.3 (2010).
\bibitem[\protect\citeauthoryear{Casares et al.}{1992}]{casares92}
J. Casares, P.A. Charles, T. Naylor, \mbox{Nature} \textbf{355}, 614 (1992).
\bibitem[\protect\citeauthoryear{Casares \& Charles}{1994}]{casares94}
J. Casares, P.A. Charles, \mbox{MNRAS} \textbf{271}, l5 (1994).
\bibitem[\protect\citeauthoryear{Casares et al.}{1995a}]{casares95a}
J. Casares, A.C. Martin, P.A. Charles, E.L. Mart\'\i{}n, R. Rebolo, E.T. Harlaftis, A.J. Castro-Tirado, 
\mbox{MNRAS} \textbf{276}, L35 (1995a).
\bibitem[\protect\citeauthoryear{Casares et al.}{1995b}]{casares95b}
J. Casares, P.A. Charles, T.R. Marsh, \mbox{MNRAS} \textbf{277}, L45 (1995b).
\bibitem[\protect\citeauthoryear{Casares et al.}{1997}]{casares97}
J. Casares, E.L. Mart\'\i{}n, P.A. Charles, P. Molaro, R. Rebolo, \mbox{NewA} \textbf{1}, 299 (1997).
\bibitem[\protect\citeauthoryear{Casares et al.}{2004}]{casares04}
J. Casares, C. Zurita, T. Shahbaz, P.A. Charles, R.P. Fender, \mbox{ApJ} \textbf{613}, L133 (2004).
\bibitem[\protect\citeauthoryear{Casares et al.}{2009}]{casares09}
J. Casares et al., \mbox{ApJS} \textbf{181}, 238 (2009).
\bibitem[\protect\citeauthoryear{Charles \& Coe}{2006}]{charles06}
P.A. Charles, M.J. Coe, \mbox{Compact stellar X-ray sources} \textbf{39}, p.215 (2006).
\bibitem[\protect\citeauthoryear{Cherepashchuk et al.}{1973}]{cherepashchuk73}
A.M. Cherepashchuk, V.M. Lyutiy, R.A. Sunyaev, \mbox{Astr.Zh.} \textbf{50}, 3 (1973).

\bibitem[\protect\citeauthoryear{{Chomiuk} et~al.}{2013}]{2013arXiv1306.6624C}
\begin{botherref}
\oauthor{\binits{L.} \bsnm{{Chomiuk}}},
\oauthor{\binits{J.} \bsnm{{Strader}}},
\oauthor{\binits{T.J.} \bsnm{{Maccarone}}},
\oauthor{\binits{J.C.A.} \bsnm{{Miller-Jones}}},
\oauthor{\binits{C.} \bsnm{{Heinke}}},
\oauthor{\binits{E.} \bsnm{{Noyola}}},
\oauthor{\binits{A.C.} \bsnm{{Seth}}},
\oauthor{\binits{S.} \bsnm{{Ransom}}},
arXiv1306.6624 (2013)
\end{botherref}
\endbibitem


\bibitem[\protect\citeauthoryear{Colbert \& Mushotzky}{1999}]{1999ApJ...519...89C}
E.~J.~M. Colbert \& R.~F. Mushotzky, \mbox{ApJ} \textbf{519} 89
(1999).

\bibitem[\protect\citeauthoryear{{Copperwheat}
  et~al.}{2005}]{2005MNRAS.362...79C}
\begin{barticle}
\bauthor{\binits{C.} \bsnm{{Copperwheat}}},
\bauthor{\binits{M.} \bsnm{{Cropper}}},
\bauthor{\binits{R.} \bsnm{{Soria}}},
\bauthor{\binits{K.} \bsnm{{Wu}}},
\bjtitle{\mnras}
\bvolume{362},
\bfpage{79}--\blpage{88}
(\byear{2005}).
\end{barticle}
\endbibitem

\bibitem[\protect\citeauthoryear{{Copperwheat}
  et~al.}{2007}]{2007MNRAS.376.1407C}
\begin{barticle}
\bauthor{\binits{C.} \bsnm{{Copperwheat}}},
\bauthor{\binits{M.} \bsnm{{Cropper}}},
\bauthor{\binits{R.} \bsnm{{Soria}}},
\bauthor{\binits{K.} \bsnm{{Wu}}},
\bjtitle{\mnras}
\bvolume{376},
\bfpage{1407}--\blpage{1423}
(\byear{2007}).
\end{barticle}
\endbibitem



\bibitem[\protect\citeauthoryear{Coriat et al.}{2012}]{coriat12}
M. Coriat, R.P. Fender, G. Dubus, \mbox{MNRAS} \textbf{424}, 1991 (2012).
\bibitem[\protect\citeauthoryear{Corral-Santana et al.}{2011}]{corral11}
J.M. Corral-Santana, J. Casares, T. Shabaz, C. Zurita, I.G. Martínez-Pais, 
P. Rodríguez-Gil,  \mbox{MNRAS} \textbf{413}, L15 (2011).
\bibitem[\protect\citeauthoryear{Corral-Santana et al.}{2013}]{corral13}
J.M. Corral-Santana, J. Casares, T. Mu\~noz-Darias, P. Rodríguez-Gil, 
T. Shabaz, C. Zurita, M.A.P. Torres, A. Tyndall, \mbox{Science} \textbf{339},
1048 (2013).
\bibitem[\protect\citeauthoryear{Cowley et al.}{1983}]{cowley83}
A. Cowley, D. Crampton, G. J.B. Hutchings, R. Remillard, J.E. Penfold, \mbox{ApJ} \textbf{272}, 118 (1983).
\bibitem[\protect\citeauthoryear{Crowther et al.}{2010}]{crowther10}
P.A. Crowther, R. Barnard, G. S. Carpano, J.S. Clark, V.S. Dhillon, A.M.T. Pollock, \mbox{MNRAS} \textbf{403}, L41 (2010).
\bibitem[\protect\citeauthoryear{Degenaar \& Wijnands}{2010}]{degenaar10}
N. Degenaar, R. Wijnands, \mbox{A\&A} \textbf{524}, 69 (2010).
\bibitem[\protect\citeauthoryear{Dhillon et al.}{2007}]{dhillon07}
V.S. Dhillon et al., \mbox{MNRAS} \textbf{378}, 825 (2007).

\bibitem[\protect\citeauthoryear{{Djorgovski} and
  {King}}{1986}]{1986ApJ...305L..61D}
\begin{barticle}
\bauthor{\binits{S.} \bsnm{{Djorgovski}}},
\bauthor{\binits{I.R.} \bsnm{{King}}},
\bjtitle{\apjl}
\bvolume{305},
\bfpage{61}--\blpage{65}
(\byear{1986}).
\end{barticle}
\endbibitem

\bibitem[\protect\citeauthoryear{{Falcke} et~al.}{2004}]{2004A&A...414..895F}
\begin{barticle}
\bauthor{\binits{H.} \bsnm{{Falcke}}},
\bauthor{\binits{E.} \bsnm{{K{\"o}rding}}},
\bauthor{\binits{S.} \bsnm{{Markoff}}},
\bjtitle{\aap}
\bvolume{414},
\bfpage{895}--\blpage{903}
(\byear{2004}).
\end{barticle}
\endbibitem

\bibitem[\protect\citeauthoryear{Farr et al.}{2011}]{farr11}
W.M. Farr, N. Sravan, A. Cantrell, L. Kreidberg, C.D. Bailyn, I. Mandel V.
Kalogera, \mbox{ApJ} \textbf{741}, 103 (2011).

\bibitem[\protect\citeauthoryear{{Farrell} et~al.}{2009}]{2009Natur.460...73F}
\begin{barticle}
\bauthor{\binits{S.A.} \bsnm{{Farrell}}},
\bauthor{\binits{N.A.} \bsnm{{Webb}}},
\bauthor{\binits{D.} \bsnm{{Barret}}},
\bauthor{\binits{O.} \bsnm{{Godet}}},
\bauthor{\binits{J.M.} \bsnm{{Rodrigues}}},
\bjtitle{\nat}
\bvolume{460},
\bfpage{73}--\blpage{75}
(\byear{2009}).
\end{barticle}
\endbibitem


\bibitem[\protect\citeauthoryear{Fender et al.}{1999}]{fender99}
R.P. Fender et al., \mbox{MNRAS} \textbf{304}, 865 (1999).

\bibitem[\protect\citeauthoryear{{Ferrarese} and
  {Merritt}}{2000}]{2000ApJ...539L...9F}
\begin{barticle}
\bauthor{\binits{L.} \bsnm{{Ferrarese}}},
\bauthor{\binits{D.} \bsnm{{Merritt}}},
\bjtitle{\apjl}
\bvolume{539},
\bfpage{9}--\blpage{12}
(\byear{2000}).
\end{barticle}
\endbibitem


\bibitem[\protect\citeauthoryear{Filippenko et al.}{1995a}]{filippenko95a}
A.V. Filippenko, T. Matheson, L.C. Ho, \mbox{ApJ} \textbf{455}, 614 (1995a).
\bibitem[\protect\citeauthoryear{Filippenko et al.}{1995b}]{filippenko95b}
A.V. Filippenko, T. Matheson, A.J. Barth, \mbox{ApJ} \textbf{455}, L139 (1995b).
\bibitem[\protect\citeauthoryear{Filippenko et al.}{1997}]{filippenko97}
A.V. Filippenko, T. Matheson, D.C. Leonard, A.J. Barth, S.D. van Dyk, \mbox{PASP} \textbf{109}, 461 (1997).
\bibitem[\protect\citeauthoryear{Filippenko et al.}{1999}]{filippenko99}
A.V. Filippenko, D.C. Leonard, T. Matheson, W. Li, E.C. Moran, A.G. Riess, \mbox{PASP} \textbf{111}, 969 (1999).
\bibitem[\protect\citeauthoryear{Filippenko \& Chornock}{2001}]{filippenko01}
A.V. Filippenko, R. Chornock, \mbox{IAUC} \textbf{7644}, 1 (2001).
\bibitem[\protect\citeauthoryear{Froning \& Robinson}{2001}]{froning01}
C.S. Froning, E.L. Robinson, \mbox{AJ} \textbf{121}, 2212 (2001).
\bibitem[\protect\citeauthoryear{Fryer \& Kalogera}{2001}]{fryer01}
C.L. Fryer, V. Kalogera, \mbox{ApJ} \textbf{554}, 548 (2001).
\bibitem[\protect\citeauthoryear{Fryer et al.}{2012}]{fryer12}
C.L. Fryer, K. Belczynski, G. Wiktorowicz, M. Dominik, V. Kalogera, D.E. Holz, 
\mbox{ApJ} \textbf{749}, 91 (2012).
\bibitem[\protect\citeauthoryear{Garcia et al.}{1996}]{garcia96} 
M.R. Garcia, P.J. Callanan, J.E. McClintock, P. Zhao, \mbox{ApJ}
\textbf{460}, 932 (1996).

\bibitem[\protect\citeauthoryear{{Gebhardt} et~al.}{2002}]{2002ApJ...578L..41G}
\begin{barticle}
\bauthor{\binits{K.} \bsnm{{Gebhardt}}},
\bauthor{\binits{R.M.} \bsnm{{Rich}}},
\bauthor{\binits{L.C.} \bsnm{{Ho}}},
\bjtitle{\apjl}
\bvolume{578},
\bfpage{41}--\blpage{45}
(\byear{2002}).
\end{barticle}
\endbibitem

\bibitem[\protect\citeauthoryear{{Gebhardt} et~al.}{2005}]{2005ApJ...634.1093G}
\begin{barticle}
\bauthor{\binits{K.} \bsnm{{Gebhardt}}},
\bauthor{\binits{R.M.} \bsnm{{Rich}}},
\bauthor{\binits{L.C.} \bsnm{{Ho}}},
\bjtitle{\apj}
\bvolume{634},
\bfpage{1093}--\blpage{1102}
(\byear{2005}).
\end{barticle}
\endbibitem

\bibitem[\protect\citeauthoryear{{Grabhorn} et~al.}{1992}]{1992ApJ...392...86G}
\begin{barticle}
\bauthor{\binits{R.P.} \bsnm{{Grabhorn}}},
\bauthor{\binits{H.N.} \bsnm{{Cohn}}},
\bauthor{\binits{P.M.} \bsnm{{Lugger}}},
\bauthor{\binits{B.W.} \bsnm{{Murphy}}},
\bjtitle{\apj}
\bvolume{392},
\bfpage{86}--\blpage{98}
(\byear{1992}).
\end{barticle}
\endbibitem


\bibitem[\protect\citeauthoryear{Gray}{1992}]{gray92}
D.F.Gray, The Observation and Analysis of Stellar Photospheres. 
CUP, Cambridge (1992).



\bibitem[\protect\citeauthoryear{Gelino et al.}{2001a}]{gelino01a} 
D.M. Gelino, T.E. Harrison, B.J. McNamara, \mbox{ApJ} \textbf{122}, 971 (2001a).
\bibitem[\protect\citeauthoryear{Gelino et al.}{2001b}]{gelino01b} 
D.M. Gelino, T.E. Harrison, J.A. Orosz, \mbox{AJ} \textbf{122}, 2668 (2001b).
\bibitem[\protect\citeauthoryear{Gelino \& Harrison}{2003}]{gelino03} 
D.M. Gelino, T.E. Harrison, \mbox{ApJ} \textbf{599}, 1254 (2003).
\bibitem[\protect\citeauthoryear{Gelino et al.}{2006}]{gelino06} 
D.M. Gelino, S. Balman, U. Kiziloglu, A. Yilmaz, E. Kalemci, J.A. Tomsick, \mbox{ApJ} \textbf{642}, 438 (2006).
\bibitem[\protect\citeauthoryear{Gies \& Bolton}{1982}]{gies82} 
D.R. Gies, C.T. Bolton, \mbox{ApJ} \textbf{260}, 240 (1982).
\bibitem[\protect\citeauthoryear{Gies \& Bolton}{1986}]{gies86} 
D.R. Gies, C.T. Bolton, \mbox{ApJ} \textbf{304}, 371 (1986).
\bibitem[\protect\citeauthoryear{Gonz\'alez Hern\'andez et al.}{2008a}]{gonzalez08a} 
J.I. Gonz\'alez Hern\'andez et al., \mbox{ApJ} \textbf{679}, 732 (2008a).
\bibitem[\protect\citeauthoryear{Gonz\'alez Hern\'andez et al.}{2008b}]{gonzalez08b} 
J.I. Gonz\'alez Hern\'andez, R. Rebolo, G. Israelian, \mbox{A\&A} \textbf{478}, 203 (2008b).
\bibitem[\protect\citeauthoryear{Gonz\'alez Hern\'andez \& Casares}{2010}]{gonzalez10} 
J.I. Gonz\'alez Hern\'andez, J. Casares, \mbox{A\&A} \textbf{516}, A58 (2010).
\bibitem[\protect\citeauthoryear{Gonz\'alez Hern\'andez et al.}{2012}]{gonzalez12} 
J.I. Gonz\'alez Hern\'andez, R. Rebolo, J. Casares, \mbox{ApJ} \textbf{744}, L25 (2012).
\bibitem[\protect\citeauthoryear{Greene et al.}{2001}]{greene01} 
J. Greene, C.D. Bailyn, J.A. Orosz, \mbox{ApJ} \textbf{554}, 1290
(2001).
\bibitem[\protect\citeauthoryear{Greene \& Ho}{2007}]{2007ApJ...670...92G}
J.~E. Greene, L.~C. Ho, \mbox{ApJ} \textbf{670}, 92 (2007).
\bibitem[\protect\citeauthoryear{Greiner et al.}{2001}]{greiner01} 
J. Greiner, J.G. Cuby, M.J. McCaughream, \mbox{Nature} \textbf{414},
522 (2001).

\bibitem[\protect\citeauthoryear{{Greiner} et~al.}{2001}]{2001A&A...373L..37G}
\begin{barticle}
\bauthor{\binits{J.} \bsnm{{Greiner}}},
\bauthor{\binits{J.G.} \bsnm{{Cuby}}},
\bauthor{\binits{M.J.} \bsnm{{McCaughrean}}},
\bauthor{\binits{A.J.} \bsnm{{Castro-Tirado}}},
\bauthor{\binits{R.E.} \bsnm{{Mennickent}}},
\bjtitle{\aap}
\bvolume{373},
\bfpage{37}--\blpage{40}
(\byear{2001}).
\end{barticle}
\endbibitem
\bibitem[\protect\citeauthoryear{Gris{\'e} et al.}{2012}]{2012ApJ...745..123G}
F. Gris{\'e}, P. Kaaret, S. Corbel, H. Feng, D. Cseh, L. Tao \mbox{APJ} \textbf{745}, 123 (2012).
\bibitem[\protect\citeauthoryear{Harlaftis et al.}{1996}]{harlaftis96}
E.T. Harlaftis, K. Horne, A.V. Filippenko, \mbox{PASP} \textbf{108}, 762 (1996).
\bibitem[\protect\citeauthoryear{Harlaftis et al.}{1997}]{harlaftis97}
E.T. Harlaftis, D. Steeghs, K. Horne, A.V. Filippenko, \mbox{AJ} \textbf{114}, 1170 (1997).
\bibitem[\protect\citeauthoryear{Harlaftis et al.}{1999}]{harlaftis99}
E.T. Harlaftis, S. Collier, K. Horne, A.V. Filippenko, \mbox{A\&A} \textbf{341}, 491 (1999).
\bibitem[\protect\citeauthoryear{Harlaftis \& Greiner}{2004}]{harlaftis04}
E.T. Harlaftis, J. Greiner, \mbox{A\&A} \textbf{414}, L13 (2004).
\bibitem[\protect\citeauthoryear{Haswell et al.}{1993}]{haswell93}
C.A. Haswell, E.L. Robinson, K. Horne, R.F. Stiening, T.M.C. Abbott, \mbox{ApJ} \textbf{411}, 802 (1993).
\bibitem[\protect\citeauthoryear{Herrero et al.}{1995}]{herrero95}
A. Herrero, R.P. Kudritzki, R. Gabler, J.M. Vilchez, A. Gabler, \mbox{A\&A} \textbf{297}, 556 (1995).
\bibitem[\protect\citeauthoryear{Hjellming \& Ruppen}{1995}]{hjellming95}
R.M. Hjellming, M.P. Rupen, \mbox{Nature} \textbf{375}, 464 (1995).
\bibitem[\protect\citeauthoryear{Hurley et al.}{2013}]{hurley13}
D.J. Hurley, P.J. Callanan, P. Elebert, M.T. Reynolds, \mbox{MNRAS} \textbf{430}, 1832 (2013).
\bibitem[\protect\citeauthoryear{Hutchings et al.}{1983}]{hutchings83}
J.B. Hutchings, D. Crampton, A. Cowley, \mbox{ApJ} \textbf{275}, L43 (1983).
\bibitem[\protect\citeauthoryear{Hutchings et al.}{1987}]{hutchings87}
J.B. Hutchings, D. Crampton, A. Cowley, L. Bianchi, I.B. Thompson, \mbox{AJ} \textbf{94}, 340 (1987).
\bibitem[\protect\citeauthoryear{Hynes et al.}{2003a}]{hynes03a}
 R.I. Hynes, D. Steeghs, J. Casares, P.A. Charles, K. O'Brien, \mbox{ApJ} \textbf{583}, L95 (2003a).
\bibitem[\protect\citeauthoryear{Hynes et al.}{2003b}]{hynes03b}
 R.I. Hynes,  P.A. Charles, J. Casares, C.A. Haswell, C. Zurita, T. Shahbaz, \mbox{MNRAS} \textbf{340}, 447 (2003b).
\bibitem[\protect\citeauthoryear{Hynes et al.}{2004}]{hynes04}
R.I. Hynes et al., \mbox{ApJ} \textbf{611}, L125 (2004).
\bibitem[\protect\citeauthoryear{Hynes}{2005}]{hynes05}
R.I. Hynes, \mbox{ASP Conf. Ser.} \textbf{330}, p.237 (2005).
\bibitem[\protect\citeauthoryear{Ioannou et al.}{2004}]{ioannou04}
Z. Ioannou, E.L. Robinson, W.F. Welsh, C.A. Haswell, \mbox{ApJ} \textbf{127}, 481 (2004).

\bibitem[\protect\citeauthoryear{{Jonker} and
  {Nelemans}}{2004}]{2004MNRAS.354..355J}
\begin{barticle}
\bauthor{\binits{P.G.} \bsnm{{Jonker}}},
\bauthor{\binits{G.} \bsnm{{Nelemans}}},
\bjtitle{\mnras}
\bvolume{354},
\bfpage{355}--\blpage{366}
(\byear{2004}).
\end{barticle}
\endbibitem


\bibitem[\protect\citeauthoryear{Jonker et al.}{2011}]{jonker11}
P.~G. Jonker et al., \mbox{ApJS} \textbf{194}, 18 (2011).

\bibitem[\protect\citeauthoryear{{Jonker} et~al.}{2012}]{2012arXiv1208.4502J}
\begin{botherref}
\oauthor{\binits{P.G.} \bsnm{{Jonker}}},
\oauthor{\binits{M.} \bsnm{{Heida}}},
\oauthor{\binits{M.A.P.} \bsnm{{Torres}}},
\oauthor{\binits{J.C.A.} \bsnm{{Miller-Jones}}},
\oauthor{\binits{A.C.} \bsnm{{Fabian}}},
\oauthor{\binits{E.M.} \bsnm{{Ratti}}},
\oauthor{\binits{G.} \bsnm{{Miniutti}}},
\oauthor{\binits{D.J.} \bsnm{{Walton}}},
\oauthor{\binits{T.P.} \bsnm{{Roberts}}},
\mbox{ApJ} \textbf{758}, 28 (2012)
\end{botherref}
\endbibitem


\bibitem[\protect\citeauthoryear{Kendrew et al.}{2012}]{kendrew12}
S. Kendrew et al., \mbox{SPIE} \textbf{8446}, 7 (2012).
\bibitem[\protect\citeauthoryear{Khargharia et al.}{2010}]{khargharia10}
J. Khargharia, C.S. Froning, E.L. Robinson, \mbox{ApJ} \textbf{716}, 1105 (2010).
\bibitem[\protect\citeauthoryear{Khargharia et al.}{2013}]{khargharia13}
J. Khargharia, C.S. Froning, E.L. Robinson, D.M. Gelino \mbox{AJ} \textbf{145}, 21 (2013).
\bibitem[\protect\citeauthoryear{Kiel \& Hurley}{ 2006}]{kiel06} 
P.D. Kiel, J.R. Hurley, \mbox{MNRAS} \textbf{ 369}, 1152 (2006).
\bibitem[\protect\citeauthoryear{King et al.}{1997}]{king97}
A.R. King, U. Kolb, E. Szuszkiewicz \mbox{ApJ} \textbf{488}, 89
(1997).

\bibitem[\protect\citeauthoryear{{Kong} et~al.}{2010}]{2010MNRAS.407L..84K}
\begin{barticle}
\bauthor{\binits{A.K.H.} \bsnm{{Kong}}},
\bauthor{\binits{C.O.} \bsnm{{Heinke}}},
\bauthor{\binits{R.} \bsnm{{di Stefano}}},
\bauthor{\binits{H.N.} \bsnm{{Cohn}}},
\bauthor{\binits{P.M.} \bsnm{{Lugger}}},
\bauthor{\binits{P.} \bsnm{{Barmby}}},
\bauthor{\binits{W.H.G.} \bsnm{{Lewin}}},
\bauthor{\binits{F.A.} \bsnm{{Primini}}},
\bjtitle{\mnras}
\bvolume{407},
\bfpage{84}--\blpage{88}
(\byear{2010}).
\end{barticle}
\endbibitem

\bibitem[\protect\citeauthoryear{Lasota}{2001}]{lasota01}
	J.-P. Lasota, \mbox{NewAR} \textbf{45} 449 (2001).
\bibitem[\protect\citeauthoryear{K{\"o}rding, Falcke \&
    Markoff}{2002}]{2002A&A...382L..13K}
E. K{\"o}rding, H. Falcke \& S. Markoff, \mbox{A\&A} \textbf{382} L31 (2002).
\bibitem[\protect\citeauthoryear{Kreidberg et al.}{2012}]{kreidberg12}
L. Kreidberg, C.D. Bailyn, W. Farr, V. Kalogera, \mbox{ApJ} \textbf{757}, 36 (2012).
\bibitem[\protect\citeauthoryear{Lattimer}{2012}]{lattimer12}
J.M. Lattimer \mbox{ARNPS} \textbf{62}, 485 (2012).
\bibitem[\protect\citeauthoryear{Larson \& Schulman}{1997}]{larson97}
D.T. Larson, E. Schulman, \mbox{AJ} \textbf{113}, 618 (1997).
\bibitem[\protect\citeauthoryear{Lestrade et al.}{1999}]{lestrade99}
J.-F. Lestrade et al., \mbox{A\&A} \textbf{344}, 1014 (1999).

\bibitem[\protect\citeauthoryear{{Liu}}{2009}]{2009ApJ...704.1628L}
\begin{barticle}
\bauthor{\binits{J.} \bsnm{{Liu}}},
\bjtitle{\apj}
\bvolume{704},
\bfpage{1628}--\blpage{1639}
(\byear{2009}).
\end{barticle}
\endbibitem


\bibitem[\protect\citeauthoryear{{Liu} et~al.}{2004}]{2004ApJ...602..249L}
\begin{barticle}
\bauthor{\binits{J.-F.} \bsnm{{Liu}}},
\bauthor{\binits{J.N.} \bsnm{{Bregman}}},
\bauthor{\binits{P.} \bsnm{{Seitzer}}},
\bjtitle{\apj}
\bvolume{602},
\bfpage{249}--\blpage{256}
(\byear{2004}).
\end{barticle}
\endbibitem

\bibitem[\protect\citeauthoryear{{Liu} et~al.}{2012}]{2012ApJ...745...89L}
\begin{barticle}
\bauthor{\binits{J.} \bsnm{{Liu}}},
\bauthor{\binits{J.} \bsnm{{Orosz}}},
\bauthor{\binits{J.N.} \bsnm{{Bregman}}},
\bjtitle{\apj}
\bvolume{745},
\bfpage{89}
(\byear{2012}).
\end{barticle}
\endbibitem


\bibitem[\protect\citeauthoryear{Lucy}{1967}]{lucy67}
L.B. Lucy, \mbox{Z.f.Astroph.} \textbf{65}, 89 (1967).

\bibitem[\protect\citeauthoryear{{L{\"u}tzgendorf}
  et~al.}{2013}]{2013A&A...555A..26L}
\begin{barticle}
\bauthor{\binits{N.} \bsnm{{L{\"u}tzgendorf}}},
\bauthor{\binits{M.} \bsnm{{Kissler-Patig}}},
\bauthor{\binits{N.} \bsnm{{Neumayer}}},
\bauthor{\binits{H.} \bsnm{{Baumgardt}}},
\bauthor{\binits{E.} \bsnm{{Noyola}}},
\bauthor{\binits{P.T.} \bsnm{{de Zeeuw}}},
\bauthor{\binits{K.} \bsnm{{Gebhardt}}},
\bauthor{\binits{B.} \bsnm{{Jalali}}},
\bauthor{\binits{A.} \bsnm{{Feldmeier}}},
\bjtitle{\aap}
\bvolume{555},
\bfpage{26}
(\byear{2013}).
\end{barticle}
\endbibitem


\bibitem[\protect\citeauthoryear{Madau \& Rees}{2001}]{2001ApJ...551L..27M}
P. Madau, M.~J. Rees, \mbox{ApJ} \textbf{551}, L27 (2001).
\bibitem[\protect\citeauthoryear{Marsh, Robinson \& Wood}{1994}]{marsh94}
T.R. Marsh, E.L. Robinson, J.H. Wood, \mbox{MNRAS} \textbf{266}, 137 (1994).
\bibitem[\protect\citeauthoryear{Martin et al.}{1995}]{martin95}
A.C. Martin, J. Casares, P.A. Charles, F. van dr Hooft, J. van Paradijs, \mbox{MNRAS} \textbf{274}, L46 (1995).
\bibitem[\protect\citeauthoryear{McClintock \& Remillard}{1986}]{mcclintock86}
J.E. McClintock, R.A. Remillard, \mbox{ApJ} \textbf{308}, 110 (1986).
\bibitem[\protect\citeauthoryear{McClintock et al.}{2001}]{mcclintock01}
J.E. McClintock, M.R. Garcia, N. Caldwell, E.E. Falco, P.M. Garnavich, P. Zhao, \mbox{ApJ} \textbf{551}, L147 (2001).
\bibitem[\protect\citeauthoryear{McClintock \& Remillard}{2006}]{mcclintock06}
J.E. McClintock, R.A. Remillard, \mbox{Black Hole Binaries}
\textbf{chap.4}, pp.157-213 (2006).

\bibitem[\protect\citeauthoryear{{Merloni} et~al.}{2003}]{2003MNRAS.345.1057M}
\begin{barticle}
\bauthor{\binits{A.} \bsnm{{Merloni}}},
\bauthor{\binits{S.} \bsnm{{Heinz}}},
\bauthor{\binits{T.} \bsnm{{di Matteo}}},
\bjtitle{\mnras}
\bvolume{345},
\bfpage{1057}--\blpage{1076}
(\byear{2003}).
\end{barticle}
\endbibitem


\bibitem[\protect\citeauthoryear{Miller-Jones et al.}{2009}]{miller-jones09}
J.C.A. Miller-Jones, P.G. Jonker, V. Dhawan, W. Brisken, M.P. Rupen, G. Nelemans, E. Gallo, 
\mbox{ApJ} \textbf{706}, L230 (2009). 

\bibitem[\protect\citeauthoryear{{Miller-Jones}
  et~al.}{2012}]{2012ApJ...755L...1M}
\begin{barticle}
\bauthor{\binits{J.C.A.} \bsnm{{Miller-Jones}}},
\bauthor{\binits{J.M.} \bsnm{{Wrobel}}},
\bauthor{\binits{G.R.} \bsnm{{Sivakoff}}},
\bauthor{\binits{C.O.} \bsnm{{Heinke}}},
\bauthor{\binits{R.E.} \bsnm{{Miller}}},
\bauthor{\binits{R.M.} \bsnm{{Plotkin}}},
\bauthor{\binits{R.} \bsnm{{Di Stefano}}},
\bauthor{\binits{J.E.} \bsnm{{Greene}}},
\bauthor{\binits{L.C.} \bsnm{{Ho}}},
\bauthor{\binits{T.D.} \bsnm{{Joseph}}},
\bauthor{\binits{A.K.H.} \bsnm{{Kong}}},
\bauthor{\binits{T.J.} \bsnm{{Maccarone}}},
\bjtitle{\apjl}
\bvolume{755},
\bfpage{1}
(\byear{2012}).
\end{barticle}
\endbibitem


\bibitem[\protect\citeauthoryear{Miller \& Hamilton}{2002}]{2002MNRAS.330..232C}
M.~C. Miller, D.~P. Hamilton, \mbox{MNRAS} \textbf{330}, 232 (2002).
\bibitem[\protect\citeauthoryear{Mineshige \& Wheeler}{1989}]{mineshige89}
S. Mineshige, J.C. Wheeler, \mbox{ApJ} \textbf{343}, 241 (1989). 
\bibitem[\protect\citeauthoryear{Mirabel \& Rodrigues}{2003}]{mirabel03}
I.F. Mirabel, I. Rodrigues, \mbox{Science} \textbf{300}, 1119 (2003). 
\bibitem[\protect\citeauthoryear{Mu\~noz-Darias et al.}{2005}]{munoz05}
T. Mu\~noz-Darias, J. Casares, I.G. Mart\'\i{}nez-Pais, \mbox{MNRAS} \textbf{635}, 502 (2005).
\bibitem[\protect\citeauthoryear{Mu\~noz-Darias et al.}{2007}]{munoz07}
T. Mu\~noz-Darias, I.G. Mart\'\i{}nez-Pais, J. Casares, V.S. Dhillon, T.R. Marsh, 
R. Cornelisse, D. Steeghs, P.A. Charles, \mbox{MNRAS} \textbf{379}, 1673 (2007).
\bibitem[\protect\citeauthoryear{Mu\~noz-Darias et al.}{2008a}]{munoz08a}
T. Mu\~noz-Darias, J. Casares, I.G. Mart\'\i{}nez-Pais, \mbox{MNRAS} \textbf{385}, 2205 (2008a).
\bibitem[\protect\citeauthoryear{Mu\~noz-Darias et al.}{2008b}]{munoz08b}
T. Mu\~noz-Darias et al., \mbox{AIP Conf Proc.} \textbf{984}, p.15 (2008b).
\bibitem[\protect\citeauthoryear{Narayan \& McClintock}{2005}]{narayan05}
R. Narayan, J.E. McClintock, \mbox{ApJ} \textbf{623}, 1017 (2005).
\bibitem[\protect\citeauthoryear{Neilsen et al.}{2008}]{neilsen08}
J. Neilsen, D. Steeghs, S.D. Vrtilek, \mbox{MNRAS} \textbf{384}, 849 (2008).
\bibitem[\protect\citeauthoryear{Ninkov et al.}{1987}]{ninkov87}
Z. Ninkov, G.A.H. Walker, S. Yang, \mbox{ApJ} \textbf{321}, 425 (1987).
\bibitem[\protect\citeauthoryear{O'Brien et al.}{2002}]{obrien02}
K. O'Brien, K. Horne, R.I. Hynes, W. Chen, C.A. Haswell, M.D. Still,
\mbox{MNRAS} \textbf{334}, 4260 (2002).
\bibitem[\protect\citeauthoryear{O'Donoghue \& Charles}{1996}]{odonoghue96}
D. O'Donoghue, P.A. Charles, \mbox{MNRAS} \textbf{282}, 191 (1996).
\bibitem[\protect\citeauthoryear{Orosz}{2001}]{orosz01b}
J.A. Orosz, \mbox{ATel} \textbf{67} (2001)
\bibitem[\protect\citeauthoryear{Orosz}{2003}]{orosz03}
J.A. Orosz, \mbox{IAU Symp.} \textbf{212}, p.365 (2003).
\bibitem[\protect\citeauthoryear{Orosz et al.}{1994}]{orosz94}
J.A. Orosz, C.D. Bailyn, J.E. McClintock, R.A. Remillard, C.B. Foltz, \mbox{ApJ} \textbf{436}, 848 (1994).
\bibitem[\protect\citeauthoryear{Orosz \& Bailyn}{1995}]{orosz95}
J.A. Orosz, C.D. Bailyn, \mbox{ApJ} \textbf{446}, L59 (1995).
\bibitem[\protect\citeauthoryear{Orosz et al.}{1996}]{orosz96}
J.A. Orosz, C.D. Bailyn, J.E. McClintock, R.A. Remillard, \mbox{ApJ} \textbf{468}, 380 (1996).
\bibitem[\protect\citeauthoryear{Orosz \& Bailyn}{1997}]{orosz97}
J.A. Orosz, C.D. Bailyn, \mbox{ApJ} \textbf{477}, 876 (1997).
\bibitem[\protect\citeauthoryear{Orosz et al.}{1998}]{orosz98}
J.A. Orosz, R.K. Jain, C.D. Bailyn,  J.E. McClintock, R.A. Remillard, \mbox{ApJ} \textbf{499}, 375 (1998).
\bibitem[\protect\citeauthoryear{Orosz \& Hauschildt}{2000}]{orosz00}
J.A. Orosz, P.H. Hauschildt, \mbox{A\&A} \textbf{364}, 265 (2000)
\bibitem[\protect\citeauthoryear{Orosz et al.}{2001}]{orosz01}
J.A. Orosz et al., \mbox{ApJ} \textbf{555}, 489 (2001)
\bibitem[\protect\citeauthoryear{Orosz et al.}{2002}]{orosz02}
J.A. Orosz et al., \mbox{ApJ} \textbf{568}, 845 (2002)
\bibitem[\protect\citeauthoryear{Orosz et al.}{2004}]{orosz04}
J.A. Orosz, J.E. McClintock, R.A. Remillard, S. Corbel, \mbox{ApJ} \textbf{616}, 376 (2004)
\bibitem[\protect\citeauthoryear{Orosz et al.}{2007}]{orosz07}
J.A. Orosz et al. \mbox{Nature} \textbf{449}, 872 (2007)
\bibitem[\protect\citeauthoryear{Orosz et al.}{2009}]{orosz09}
J.A. Orosz et al. \mbox{ApJ} \textbf{697}, 573 (2009)
\bibitem[\protect\citeauthoryear{Orosz et al.}{2011a}]{orosz11a}
J.A. Orosz, J.F. Steiner, J.E. McClintock, M.A.P. Torres, R.A. Remillard, C.D. Bailyn, J.M. Miller, 
\mbox{ApJ} \textbf{730}, 75 (2011a)
\bibitem[\protect\citeauthoryear{Orosz et al.}{2011b}]{orosz11b}
J.A. Orosz, J.E. McClintock, J.P. Aufdenberg, R.A. Remillard, M. Reid, 
R. Narayan, L. Gou \mbox{ApJ} \textbf{742}, 840 (2011b)

\bibitem[\protect\citeauthoryear{{\"O}zel et al.}{2010}]{ozel10}
F. {\"O}zel, D. Psaltis, R. Narayan, J.E. McClintock
\mbox{ApJ} \textbf{725}, 1918 (2010)

\bibitem[\protect\citeauthoryear{{Ohsuga} and
  {Mineshige}}{2011}]{2011ApJ...736....2O}
\begin{barticle}
\bauthor{\binits{K.} \bsnm{{Ohsuga}}},
\bauthor{\binits{S.} \bsnm{{Mineshige}}},
\bjtitle{\apj}
\bvolume{736},
\bfpage{2}
(\byear{2011}).
\end{barticle}
\endbibitem

\bibitem[\protect\citeauthoryear{{O'Leary} and
  {Loeb}}{2012}]{2012MNRAS.421.2737O}
\begin{barticle}
\bauthor{\binits{R.M.} \bsnm{{O'Leary}}},
\bauthor{\binits{A.} \bsnm{{Loeb}}},
\bjtitle{\mnras}
\bvolume{421},
\bfpage{2737}--\blpage{2750}
(\byear{2012}).
\end{barticle}
\endbibitem


\bibitem[\protect\citeauthoryear{Pavlenko et al.}{1996}]{pavlenko96}
E.P. Pavlenko, A.C. Martin, J. Casares, P.A. Charles, N.A. Ketsaris, 
\mbox{MNRAS} \textbf{281}, 1094 (1996).
\bibitem[\protect\citeauthoryear{Pietsch et al.}{2006}]{pietsch06}
W. Pietsch et al., \mbox{ApJ} \textbf{646}, 420 (2006).

\bibitem[\protect\citeauthoryear{Podsiadlowski et al.}{2003}]{podsiadlowski2003}
Ph. Podsiadlowski, S. Rappaport, Z. Han, \mbox{MNRAS} \textbf{341}, 385 (2003).

\bibitem[\protect\citeauthoryear{{Pooley} and
  {Rappaport}}{2006}]{2006ApJ...644L..45P}
\begin{barticle}
\bauthor{\binits{D.} \bsnm{{Pooley}}},
\bauthor{\binits{S.} \bsnm{{Rappaport}}},
\bjtitle{\apjl}
\bvolume{644},
\bfpage{45}--\blpage{48}
(\byear{2006}).
\end{barticle}
\endbibitem

\bibitem[\protect\citeauthoryear{{Portegies Zwart} and
  {McMillan}}{2002}]{2002ApJ...576..899P}
\begin{barticle}
\bauthor{\binits{S.F.} \bsnm{{Portegies Zwart}}},
\bauthor{\binits{S.L.W.} \bsnm{{McMillan}}},
\bjtitle{\apj}
\bvolume{576},
\bfpage{899}--\blpage{907}
(\byear{2002}).
\end{barticle}
\endbibitem


\bibitem[\protect\citeauthoryear{{Raimundo} et~al.}{2010}]{2010MNRAS.408.1714R}
\begin{barticle}
\bauthor{\binits{S.I.} \bsnm{{Raimundo}}},
\bauthor{\binits{A.C.} \bsnm{{Fabian}}},
\bauthor{\binits{F.E.} \bsnm{{Bauer}}},
\bauthor{\binits{D.M.} \bsnm{{Alexander}}},
\bauthor{\binits{W.N.} \bsnm{{Brandt}}},
\bauthor{\binits{B.} \bsnm{{Luo}}},
\bauthor{\binits{R.V.} \bsnm{{Vasudevan}}},
\bauthor{\binits{Y.Q.} \bsnm{{Xue}}},
\bjtitle{\mnras}
\bvolume{408},
\bfpage{1714}--\blpage{1720}
(\byear{2010}).
\end{barticle}
\endbibitem



\bibitem[\protect\citeauthoryear{Rappaport \&
    Joss}{1983}]{rappaport83}
S.A. Rappaport, P.C. Joss, \mbox{in Accretion Driven X-ray Sources}
\textbf{CUP}, 33 (1983).

\bibitem[\protect\citeauthoryear{{Rappaport}
  et~al.}{2005}]{2005MNRAS.356..401R}
\begin{barticle}
\bauthor{\binits{S.A.} \bsnm{{Rappaport}}},
\bauthor{\binits{P.} \bsnm{{Podsiadlowski}}},
\bauthor{\binits{E.} \bsnm{{Pfahl}}},
\bjtitle{\mnras}
\bvolume{356},
\bfpage{401}--\blpage{414}
(\byear{2005}).
\end{barticle}
\endbibitem

\bibitem[\protect\citeauthoryear{{Rashkov} and
  {Madau}}{2013}]{2013arXiv1303.3929R}
\begin{botherref}
\oauthor{\binits{V.} \bsnm{{Rashkov}}},
\oauthor{\binits{P.} \bsnm{{Madau}}},
\mbox{ApJ}, arXiv1303.3929 (2013, submitted)
\end{botherref}
\endbibitem

\bibitem[\protect\citeauthoryear{Reid et al.}{2011}]{reid11}
M.J. Reid, J.E. McClintock, R. Narayan, L. Gou, R.A. Remillard,, J.A. Orosz 
\mbox{ApJ} \textbf{742}, 83 (2011).
\bibitem[\protect\citeauthoryear{Remillard et al.}{1992}]{remillard92}
R.A. Remillard, J.E. McClintock, C.D. Bailyn, \mbox{ApJ} \textbf{399}, L145 (1992).
\bibitem[\protect\citeauthoryear{Remillard et al.}{1996}]{remillard96}
R.A. Remillard, J.A. Orosz, J.E. McClintock, C.D. Bailyn, \mbox{ApJ} \textbf{459}, 226 (1996).
\bibitem[\protect\citeauthoryear{Remillard \& McClintock}{2006}]{remillard06}
R.A. Remillard, J.E. McClintock, \mbox{ARAA} \textbf{44}, 49 (2006).
\bibitem[\protect\citeauthoryear{Reynolds et al.}{2007}]{reynolds07}
M.T. Reynolds, P.J. Callanan, A.V. Filippenko, \mbox{MNRAS}
\textbf{374}, 657 (2007).
\bibitem[\protect\citeauthoryear{Ricotti, Ostriker \& Mack}{2003}]{2008ApJ...680..829R}
M. Ricotti, J.~P. Ostriker, K.~J. Mack, \mbox{ApJ} \textbf{680} 829 (2008).

\bibitem[\protect\citeauthoryear{{Roberts} et~al.}{2008}]{2008MNRAS.387...73R}
\begin{barticle}
\bauthor{\binits{T.P.} \bsnm{{Roberts}}},
\bauthor{\binits{A.J.} \bsnm{{Levan}}},
\bauthor{\binits{M.R.} \bsnm{{Goad}}},
\bjtitle{\mnras}
\bvolume{387},
\bfpage{73}--\blpage{78}
(\byear{2008}).
\end{barticle}
\endbibitem


\bibitem[\protect\citeauthoryear{{Roberts} et~al.}{2011}]{2011AN....332..398R}
\begin{barticle}
\bauthor{\binits{T.P.} \bsnm{{Roberts}}},
\bauthor{\binits{J.C.} \bsnm{{Gladstone}}},
\bauthor{\binits{A.D.} \bsnm{{Goulding}}},
\bauthor{\binits{A.M.} \bsnm{{Swinbank}}},
\bauthor{\binits{M.J.} \bsnm{{Ward}}},
\bauthor{\binits{M.R.} \bsnm{{Goad}}},
\bauthor{\binits{A.J.} \bsnm{{Levan}}},
\bjtitle{Astronomische Nachrichten}
\bvolume{332},
\bfpage{398}
(\byear{2011}).
\end{barticle}
\endbibitem

\bibitem[\protect\citeauthoryear{Romani}{1998}]{romani98}
R.W. Romani, \mbox{A\&A} \textbf{333}, 583 (1998).
\bibitem[\protect\citeauthoryear{Ritter \& King}{2002}]{ritter02}
H. Ritter, A.R. King, \mbox{ASP} \textbf{261}, 531 (2002).
\bibitem[\protect\citeauthoryear{Sanwal et al.}{1996}]{sanwal96}
D. Sanwal et al., \mbox{ApJ} \textbf{460}, 337 (1996).
\bibitem[\protect\citeauthoryear{Sarna}{1989}]{sarna89}
M.J. Sarna, \mbox{A\&A} \textbf{224}, 98 (1989).
\bibitem[\protect\citeauthoryear{Shahbaz}{2003}]{shahbaz03a}
T. Shahbaz, \mbox{MNRAS} \textbf{339}, 1031 (2003).
\bibitem[\protect\citeauthoryear{Shahbaz et al.}{1994a}]{shahbaz94a}
T. Shahbaz, T. Naylor, P.A. Charles, \mbox{MNRAS} \textbf{268}, 756 (1994a).
\bibitem[\protect\citeauthoryear{Shahbaz et al.}{1994b}]{shahbaz94b}
T. Shahbaz, F.A. Ringwald, J.C. Bunn, T. Naylor, P.A. Charles, J. Casares, \mbox{MNRAS} \textbf{271}, L10 (1994b).
\bibitem[\protect\citeauthoryear{Shahbaz et al.}{1996}]{shahbaz96}
T. Shahbaz, F. van der Hooft, P.A. Charles, J. Casares, J. van Paradijs, \mbox{MNRAS} \textbf{282}, L47 (1996).
\bibitem[\protect\citeauthoryear{Shahbaz et al.}{1997}]{shahbaz97}
T. Shahbaz, T. Naylor, P.A. Charles,\mbox{MNRAS} \textbf{285}, 607 (1997).
\bibitem[\protect\citeauthoryear{Shahbaz et al.}{1999}]{shahbaz99}
T. Shahbaz, F. van der Hooft, J. Casares, P.A. Charles, J. van Paradijs, \mbox{MNRAS} \textbf{306}, 889 (1999).
\bibitem[\protect\citeauthoryear{Shahbaz et al.}{2003}]{shahbaz03b}
T. Shahbaz, V.S. Dhillon, T.R. Marsh, C. Zurita, C.A. Haswell, P.A. Charles,
R.I. Hynes, J. Casares, \mbox{MNRAS} \textbf{346}, 1116 (2003).
\bibitem[\protect\citeauthoryear{Shahbaz et al.}{2010}]{shahbaz10}
T. Shahbaz, V.S. Dhillon, T.R. Marsh, J. Casares, C. Zurita, P.A. Charles, 
\mbox{MNRAS} \textbf{403}, 2167 (2010).
\bibitem[\protect\citeauthoryear{Shahbaz et al.}{2013}]{shahbaz13}
T. Shahbaz, D.M. Russell, C. Zurita, J. Casares, J.M. Corral-Santana, V.S. Dhillon, T.R. Marsh, 
\mbox{MNRAS} \textbf{434}, 2696 (2013).

\bibitem[\protect\citeauthoryear{Shaposhnikov \& Titarchuk}{2007}]{shaposhnikov07}
N. Shaposhnikov, L. Titarchuk, \mbox{ApJ} \textbf{663}, 445 (2007).

\bibitem[\protect\citeauthoryear{Silverman \& Filippenko}{2008}]{silver08}
J.M. Silverman, A.V. Filippenko, \mbox{ApJ} \textbf{678}, L17 (2008).


\bibitem[\protect\citeauthoryear{Song et al.}{2010}]{song10}
L. Song et al., \mbox{AJ} \textbf{140}, 794 (2010).

\bibitem[\protect\citeauthoryear{{Soria} et~al.}{2012}]{2012ApJ...750..152S}
\begin{barticle}
\bauthor{\binits{R.} \bsnm{{Soria}}},
\bauthor{\binits{K.D.} \bsnm{{Kuntz}}},
\bauthor{\binits{P.F.} \bsnm{{Winkler}}},
\bauthor{\binits{W.P.} \bsnm{{Blair}}},
\bauthor{\binits{K.S.} \bsnm{{Long}}},
\bauthor{\binits{P.P.} \bsnm{{Plucinsky}}},
\bauthor{\binits{B.C.} \bsnm{{Whitmore}}},
\bjtitle{\apj}
\bvolume{750},
\bfpage{152}
(\byear{2012}).
\end{barticle}
\endbibitem


\bibitem[\protect\citeauthoryear{Steeghs \& Casares}{2002}]{steeghs02}
D. Steeghs, J. Casares, \mbox{ApJ} \textbf{568}, 273 (2002).
\bibitem[\protect\citeauthoryear{Steeghs et al.}{2013}]{steeghs13}
D. Steeghs, J.E. McClintock, S.G. Parsons, M.J. Reid, S. Littlefair,
V.S. Dhillon, \mbox{ApJ} \textbf{768}, 185 (2013).

\bibitem[\protect\citeauthoryear{{Strader} et~al.}{2012a}]{2012ApJ...750L..27S}
\begin{barticle}
\bauthor{\binits{J.} \bsnm{{Strader}}},
\bauthor{\binits{L.} \bsnm{{Chomiuk}}},
\bauthor{\binits{T.J.} \bsnm{{Maccarone}}},
\bauthor{\binits{J.C.A.} \bsnm{{Miller-Jones}}},
\bauthor{\binits{A.C.} \bsnm{{Seth}}},
\bauthor{\binits{C.O.} \bsnm{{Heinke}}},
\bauthor{\binits{G.R.} \bsnm{{Sivakoff}}},
\bjtitle{\apjl}
\bvolume{750},
\bfpage{27}
(\byear{2012}a).
\end{barticle}
\endbibitem

\bibitem[\protect\citeauthoryear{{Strader} et~al.}{2012b}]{2012Natur.490...71S}
\begin{barticle}
\bauthor{\binits{J.} \bsnm{{Strader}}},
\bauthor{\binits{L.} \bsnm{{Chomiuk}}},
\bauthor{\binits{T.J.} \bsnm{{Maccarone}}},
\bauthor{\binits{J.C.A.} \bsnm{{Miller-Jones}}},
\bauthor{\binits{A.C.} \bsnm{{Seth}}},
\bjtitle{\nat}
\bvolume{490},
\bfpage{71}--\blpage{73}
(\byear{2012}b).
\end{barticle}
\endbibitem



\bibitem[\protect\citeauthoryear{{Strohmayer} and
  {Mushotzky}}{2003}]{2003ApJ...586L..61S}
\begin{barticle}
\bauthor{\binits{T.E.} \bsnm{{Strohmayer}}},
\bauthor{\binits{R.F.} \bsnm{{Mushotzky}}},
\bjtitle{\apjl}
\bvolume{586},
\bfpage{61}--\blpage{64}
(\byear{2003}).
\end{barticle}
\endbibitem

\bibitem[\protect\citeauthoryear{Tanaka \& Shibazaki}{1996}]{tanaka96}
Y. Tanaka, N. Shibazaki, \mbox{ARAA} \textbf{34}, 607 (1996).
\bibitem[\protect\citeauthoryear{Torres et al.}{2004}]{torres04}
M.A.P. Torres, P.J. Callanan, M.R. Garcia, P. Zhao, S. Laycock, A.K.H. Kong, \mbox{ApJ} \textbf{612}, 1026 (2004).
\bibitem[\protect\citeauthoryear{Torres et al.}{2013}]{torres13}
M.A.P. Torres et al., \mbox{MNRAS} arXiv:1310.02241 (2013, submitted).
\bibitem[\protect\citeauthoryear{Unwin et al.}{2008}]{unwin08}
S.C. Unwin, M. Shao, S.J. Edberg, \mbox{SPIE} \textbf{7013}, 78
(2008).

\bibitem[\protect\citeauthoryear{{Ulvestad} et~al.}{2007}]{2007ApJ...661L.151U}
\begin{barticle}
\bauthor{\binits{J.S.} \bsnm{{Ulvestad}}},
\bauthor{\binits{J.E.} \bsnm{{Greene}}},
\bauthor{\binits{L.C.} \bsnm{{Ho}}},
\bjtitle{\apjl}
\bvolume{661},
\bfpage{151}--\blpage{154}
(\byear{2007}).
\end{barticle}
\endbibitem

\bibitem[\protect\citeauthoryear{{van der Marel}}{2003}]{2003AIPC..686..115V}
\begin{bchapter}
\bauthor{\binits{R.P.} \bsnm{{van der Marel}}},
in \bbtitle{The Astrophysics of Gravitational Wave Sources},
ed. by \beditor{\binits{J.M.} \bsnm{{Centrella}}}
\bsertitle{American Institute of Physics Conference Series},
vol. \bseriesno{686},
\byear{2003},
pp. \bfpage{115}--\blpage{124}.
\end{bchapter}
\endbibitem


\bibitem[\protect\citeauthoryear{Valsecchi et al.}{2010}]{valsecchi10}
F. Valsecchi et al., \mbox{Nature} \textbf{468}, 7320 (2010).
\bibitem[\protect\citeauthoryear{Val-Baker et al.}{2007}]{val-baker07}
A.K.F. Val-Baker, A.J. Norton, I. Negueruela, \mbox{AIP} \textbf{924}, 530 (2007).
\bibitem[\protect\citeauthoryear{van den Heuvel}{1992}]{vandenheuvel92}
E.P.J. van den Heuvel, \mbox{Proc. Inter. Space Year Conf. ESA ISY-3}, p.29 (1992).
\bibitem[\protect\citeauthoryear{van der Hooft et al.}{1998}]{vanderhooft98}
F. van der Hooft, M.H.M. Heemskerk, F. Alberts, J. van Paradijs,
\mbox{A\&A} \textbf{329}, 538 (1998).
\bibitem[\protect\citeauthoryear{van Paradijs \&
McClintock}{1995}]{vanparadijs95}
J. van Paradijs, J.E. McClintock, \mbox{X-ray binaries}, Cambridge Astrophysics
Series \textbf{26}, p.58 (1995).
\bibitem[\protect\citeauthoryear{Volonteri}{2010}]{2010Natur.466.1049V}
M. Volonteri, \mbox{Nature} \textbf{466} 1049 (2010).
\bibitem[\protect\citeauthoryear{Volonteri}{2012}]{2012Sci...337..544V}
M. Volonteri, \mbox{Science} \textbf{337} 544 (2012).
\bibitem[\protect\citeauthoryear{Wade \& Horne}{1988}]{wade88}
R.A. Wade, K. Horne, \mbox{ApJ} \textbf{324}, 411 (1988).
\bibitem[\protect\citeauthoryear{Wagner et al.}{1992}]{wagner92}
R.M. Wagner, T.J. Kreidl, S.B. Howell, S.G. Starrfield, \mbox{ApJ} \textbf{401}, L25 (1992).
\bibitem[\protect\citeauthoryear{Wagner et al.}{2001}]{wagner01}
R.M. Wagner, C.B. Foltz, T. Shahbaz, J. Casares, P.A. Charles,
S.G. Starrfield, P. Hewett, \mbox{ApJ} \textbf{556}, 42 (2001).
\bibitem[\protect\citeauthoryear{Walton et
    al.}{2011}]{2011MNRAS.416.1844W}
D.~J. Walton, T.~P. Roberts, S. Mateos, V. Heard, \mbox{MNRAS}
\textbf{416}, 1844 (2011).
\bibitem[\protect\citeauthoryear{Webb et al.}{2000}]{webb00}
N.A. Webb, T. Naylor, Z. Ioannou, P.A. Charles, T. Shahbaz, \mbox{MNRAS} \textbf{317}, 528 (2000).
\bibitem[\protect\citeauthoryear{Webster \& Murdin}{1972}]{webster72}
B.L. Webster, P. Murdin, \mbox{Nature} \textbf{235}, 37 (1972).
\bibitem[\protect\citeauthoryear{Wellstein \& Langer}{1999}]{wellstein99}
S. Wellstein, N. Langer, \mbox{A\&A} \textbf{350}, 148 (1999).

\bibitem[\protect\citeauthoryear{{Wiersema} et~al.}{2010}]{2010ApJ...721L.102W}
\begin{barticle}
\bauthor{\binits{K.} \bsnm{{Wiersema}}},
\bauthor{\binits{S.A.} \bsnm{{Farrell}}},
\bauthor{\binits{N.A.} \bsnm{{Webb}}},
\bauthor{\binits{M.} \bsnm{{Servillat}}},
\bauthor{\binits{T.J.} \bsnm{{Maccarone}}},
\bauthor{\binits{D.} \bsnm{{Barret}}},
\bauthor{\binits{O.} \bsnm{{Godet}}},
\bjtitle{\apjl}
\bvolume{721},
\bfpage{102}--\blpage{106}
(\byear{2010}).
\end{barticle}
\endbibitem

\bibitem[\protect\citeauthoryear{Witte \& Savonije}{2001}]{witte01}
M.G. Witte, G.J. Savonije, \mbox{A\&A} \textbf{366}, 840 (2001).
\bibitem[\protect\citeauthoryear{Wong et al.}{2012}]{wong12}
T-W. Wong, F. Valsecchi, T. Fragos, V. Kalogera, \mbox{ApJ} \textbf{747}, 111 (2012).
\bibitem[\protect\citeauthoryear{Yungelson et al.}{2006}]{yungelson06}
L.R. Yungelson, J.P. Lasota, G. Nelemans, G. Dubus, E.P.J. van den Heuvel, J. Dewi, 
S. Portegies Zwart, \mbox{A\&A} \textbf{454}, 559 (2006).
\bibitem[\protect\citeauthoryear{Zi\'olkowski}{2005}]{ziolkowski05}
J. Zi\'olkowski,  \mbox{MNRAS} \textbf{358}, 851 (2005).
\bibitem[\protect\citeauthoryear{Zurita et al.}{2002}]{zurita02}
C. Zurita,  et al.,  \mbox{MNRAS} \textbf{333}, 791 (2002).
\bibitem[\protect\citeauthoryear{Zurita et al.}{2003}]{zurita03}
C. Zurita, J. Casares, T. Shahbaz, \mbox{ApJ} \textbf{582}, 369 (2003).

\end{thebibliography}

{}

\end{document}